\documentclass[twocolumn,superscriptaddress,float,prb]{revtex4-1}
\usepackage{graphicx,amsfonts,amssymb,amsmath,hyperref,hypcap,enumerate}
\usepackage{mathrsfs}
\usepackage{color}
\usepackage{cleveref}

\newcommand{\bS}{\boldsymbol{S}}

\newcommand{\pp}{\boldsymbol{p}}
\newcommand{\qq}{\boldsymbol{q}}
\newcommand{\QQ}{\boldsymbol{Q}}

\newcommand{\rr}{\boldsymbol{r}}

\newcommand{\kk}{\boldsymbol{k}}

\newcommand{\bb}{\boldsymbol{b}}

\newcommand{\mm}{\boldsymbol{m}}

\newcommand{\SW}{\text{SW}}

\begin{document}
\title{Quantum Geometry and Stability of Moir\'e Flatband Ferromagnetism}

\author{Fengcheng Wu}
\email{wufcheng@umd.edu}
\affiliation{Condensed Matter Theory Center and Joint Quantum Institute, Department of Physics, University of Maryland, College Park, Maryland 20742, USA}
\author{S. Das Sarma}
\affiliation{Condensed Matter Theory Center and Joint Quantum Institute, Department of Physics, University of Maryland, College Park, Maryland 20742, USA}

%\date{\today}

\begin{abstract}
	Several moir\'e systems created by various twisted bilayers have manifested magnetism under flatband conditions leading to enhanced interaction effects. We theoretically study stability of moir\'e flatband ferromagnetism against collective excitations, with a focus on the effects of Bloch band quantum geometry. The spin magnon spectrum is calculated using different approaches, including Bethe-Salpeter equation, single mode approximation, and an analytical theory. One of our main results is an analytical expression for the spin stiffness in terms of the Coulomb interaction potential, the Berry curvatures, and the quantum metric tensor, where the last two quantities characterize the quantum geometry of moir\'e bands.  This analytical theory shows that Berry curvatures play an important role in stiffening the spin magnons. Furthermore, we construct an effective field theory for the magnetization fluctuations, and show explicitly that skyrmion excitations bind an integer number of electrons that is proportional to the Bloch band Chern number and the skyrmion winding number.
\end{abstract}

\maketitle

\section{Introduction}
\label{sec:intro}

Twisted bilayers with nearly flat moir\'e bands \cite{Bistritzer2011} provide a versatile platform to realize novel quantum states of matter induced by strongly enhanced many-body interaction effects. In particular, the possibility of tuning interaction by controlling the twist angle leads to a new experimental paradigm. Two prototypical interaction driven states in moir\'e flatbands are superconductors and correlated insulators \cite{Cao2018Super,Cao2018Magnetic,Dean2018tuning,lu2019superconductors}. Here we use flatbands to refer to bands with a narrow (not necessarily zero) bandwidth that is comparable to the interaction strength. While superconductivity in twisted bilayers can appear at generic  filling factors \cite{lu2019superconductors}, correlated insulators typically develop only at certain commensurate filling factors (i.e., integer number of electrons/holes per moir\'e cell). Moir\'e superconductivity represents a theoretical challenge, and various pairing mechanisms have been explored\cite{Balents2018,Liu2018chiral,Senthil2018,Wu2018phonon,wu2019phonon,sarma2020electron,Heikkila2018,Isobe2018,Lian2018twisted,khalaf2020charged}. On the other hand, the correlated insulators are generally believed to be driven by Coulomb interactions, although their exact nature in many situations remains an open question that is under active study \cite{Balents2018,Liu2018chiral,Senthil2018,Koshino2018,Kang2018,khalaf2020charged,Rademaker2019,bultinck2019ground,zhang2020correlated,kang2020non,hsu2020topological,cea2020band}. One possible scenario, as proposed in several theoretical works \cite{Zhang2019,xie2018nature,kang2018strong,Seo2019,WuTITMD,LiuMulti,Wolf2019}, is flatband ferromagnetism with spin and/or valley polarization, which naturally leads to insulating states at commensurate fillings when the interaction strength is strong enough to open up a full gap at the Fermi level. Evidence of ferromagnetism has indeed been experimentally observed in many moir\'e systems, including twisted bilayer graphene aligned to hBN \cite{sharpe2019emergent,serlin2020intrinsic}, twisted double bilayer graphene \cite{Shen2020,liu2019spin,cao2020tunable,Burg2019}, ABC trilayer graphene on hBN \cite{chen2020tunable}, and twisted monolayer-bilayer graphene \cite{polshyn2020nonvolatile,chen2020electrically}.  Remarkably, quantum anomalous Hall effects have been reported in many of the above systems \cite{sharpe2019emergent,serlin2020intrinsic,chen2020tunable,polshyn2020nonvolatile}. The quantum anomalous Hall states form when the underlying moir\'e bands carry valley contrast Chern numbers and interactions generate valley polarized ferromagnets. This valley Ising ordered Chern insulator has been theoretically justified \cite{Zhang2019Twisted,bultinck2019anomalous,repellin2019ferromagnetism,alavirad2019ferromagnetism,Wu2020Collective,liu2019correlated} and its properties are under active study \cite{liu2019anomalous,He2020,zhu2020curious,su2020switching,kwan2020exciton,bomerich2020skyrmion}, while more exotic states have also been proposed for the observed anomalous Hall effects\cite{kwan2020excitonic,zhang2020quantum,stefanidis2020excitonic}. The interesting interplay of ferromagnetism, quantum geometry (e.g., Berry curvatures), and topology (e.g., Chern numbers) makes the magnetic properties of moir\'e systems theoretically intriguing and challenging.

In this paper, we study moir\'e flatband ferromagnetism, with a motivation towards a deeper understanding of its stability. A particular goal of our work is to obtain an explicit connection between quantum geometry and moir\'e flatband  ferromagnetism. Ferromagnets with maximal flavor polarization can be exact eigenstates of many-body Hamiltonians, but whether they realize the true ground state of particular interacting systems generally stands as a hard theoretical problem, with only a few known rigorous results for certain models \cite{Nagaoka1966Ferro,Lieb1989,mielke1992exact}. Here we consider a more tractable problem, that is, whether ferromagnets are at the local energy minima in the configuration space of many-body states, and particularly, whether ferromagnets are robust against one-magnon collective excitations. The one magnon excitations \cite{Wu2020Collective,alavirad2019ferromagnetism} refer to states with a total flavor polarization that is reduced by one quantum compared to that of the maximally polarized ferromagnets.  Given that ferromagnetism has been experimentally observed in several moir\'e systems, our work on its stability is particularly relevant.

Theoretically establishing the guaranteed existence of ground state ferromagnetism in moir\'e systems (or in any system) requires very accurate knowledge of the band structure and the microscopic interaction details, and then solving the many-body problem exactly, which is beyond the scope of this theoretical work.  What we establish in this work is that such  moir\'e flatband ferromagnetism, if it exists, is closely connected with the underlying band quantum geometry.  Within a mean field theory such ferromagnetism emerges naturally in the interacting flatband system, and in the absence of other bands, the flatband ferromagnetism within the Hartree-Fock theory can be an exact solution (provided there are no first order transitions to some other unknown lower energy states) similar to what happens in quantum Hall ferromagnetism \cite{Moon1995}.  Based on this Hartree-Fock theory, we further provide a detailed analytical theoretical study of the ferromagnetic stability as well as a sharp geometric interpretation of the moir\'e flatband ferromagnetism. 

We focus on spin magnons and calculate its excitation spectrum (i.e., the spin wave energy) using a variety of approaches, including Bethe-Salpeter equation \cite{Wu2020Collective}, single mode approximation and an analytical theory. The spin wave mode is a gapless Goldstone mode because of spontaneous spin SU(2) symmetry breaking in the ferromagnet. The spin wave energy is a quadratic function of momentum in the long-wavelength limit, which can be used to extract the spin stiffness.   Our main result is an analytical expression [Eq.~\eqref{eq:rhosOVg}] for the spin stiffness $\rho_s$ in terms of three quantities, the interaction potential $V(\qq)$, the Berry curvature $\Omega_{\kk}$, and the quantum metric (also known as Fubini-Study metric) tensor $\hat{g}_{\kk}$, where the last two quantities characterize the quantum geometry of the moir\'e bands. We make two remarks about this result.  (1) $|\Omega_{\kk}|$ contributes to stiffen the spin magnons, while $\hat{g}_{\kk}$ tends to suppress $\rho_s$. It is important to note that the absolute value of Berry curvatures, i.e., $|\Omega_{\kk}|$, enters into the expression of $\rho_s$, but the sign of $\Omega_{\kk}$ does not. Therefore, a topologically trivial band with a zero Chern number but finite Berry curvatures can still support ferromagnetism. (2) The quantum metric tensor $\hat{g}_{\kk}$ and the Berry curvature $\Omega_{\kk}$ are related by an inequality $\text{Tr} \hat{g}_{\kk} \geq |\Omega_{\kk}|$, as proved in Ref.~\onlinecite{Roy2014Geometry}. After approximating $\text{Tr} \hat{g}_{\kk}$ by $|\Omega_{\kk}|$, we can express $\rho_s$ in terms of the characteristic interaction strength $e^2/(\epsilon a_M)$ ($a_M$ being the moir\'e period) and $|\Omega_{\kk}|$, as shown in Eq.~\eqref{eq:rhosVO}, with $\rho_s$ being proportional to  $\int d \kk ~|\Omega_{\kk}|^{1/2} ~$. Remarkably, we find that Eq.~\eqref{eq:rhosVO}, despite being approximate, provides a semiquantitative estimation of the spin stiffness compared to that obtained from the Bethe-Salpeter equation. This indicates that Berry curvatures play an important role in stiffening the spin magnons, and therefore, stabilizing the ferromagnetic states. Moir\'e bands in twisted bilayers that break $\hat{C}_{2z}$ symmetry (a twofold rotation around the out-of-plane $\hat{z}$ axis) generically carry large Berry curvatures, thus producing stable ferromagnetism. Our theory provides a unified picture on why ferromagnetism is commonly found in moir\'e flatbands, seemingly independent of microscopic materials details.

We discuss the connection of this work with related studies. This work is a continuation of our previous paper \cite{Wu2020Collective} where the collective excitation spectra were obtained by numerically solving the Bethe-Salpeter equation. The analytical study presented in this work can be viewed as a generalization of the ferromagnetism physics from Landau levels \cite{Moon1995} to moir\'e bands, which carry, respectively, uniform and nonuniform quantum geometry. This is of course perfectly understandable in view of the moir\'e flatband ferromagnetism being analogous to quantum Hall ferromagnetism \cite{Moon1995,Yang2006} where an isolated Landau level is known to be an interaction-driven ferromagnet. The effects of Berry curvatures on $\rho_s$ for moir\'e flatband ferromagnetism have been discussed in Refs.~\onlinecite{Zhang2019,Chatterjee2020Symm,bultinck2019anomalous,repellin2019ferromagnetism,khalaf2020charged}, but the role of quantum metric in determining $\rho_s$ has not been explicitly demonstrated previously to our knowledge. Therefore, our main results in Eqs.~\eqref{eq:rhosOVg} and ~\eqref{eq:rhosVO} are new. While our work is based on a momentum-space approach, ferromagnetism could also be studied using real-space approaches \cite{kang2018strong,Seo2019,huang2020quantum}. The effects of quantum geometry on collective excitations have been studied in other contexts, including valley excitons in two-dimensional semiconductors \cite{Srivastava2015,Zhou2015Berry} and superfluid weight in  superconductors \cite{peotta2015superfluidity,hu2019geometric,julku2019superfluid,xie2019topology}.

Our paper is organized as follows. In Sec.~\ref{sec:bandFerr}, we set up preliminaries on single-particle moir\'e band theory as well as the Hartree-Fock theory for interaction driven ferromagnetic insulators. We use twisted bilayer graphene aligned to hBN as a convenient model system, where the Chern numbers of moir\'e bands can be theoretically tuned. In Sec.~\ref{sec:spinwave}, we present our theory on spin wave energy and spin stiffness. In Sec.~\ref{sec:skyrmion}, we construct an effective Lagrangian for low-energy and long-wavelength magnetization fluctuations, which is another approach to obtain the spin wave mode. We also provide a derivation that shows skyrmion excitations bind an integer number of electrons that is proportional to the Bloch band Chern number and the skyrmion winding number. In Sec.~\ref{sec:dis}, we make a brief summary. Appendix~\ref{app} gives a proof for the inequality $\text{Tr} \hat{g}_{\kk} \geq |\Omega_{\kk}|$, and explicitly shows the connection between moir\'e and quantum Hall ferromagnetism.

\section{Moir\'e bands and Ferromagnetism}
\label{sec:bandFerr}

We use twisted bilayer graphene (TBG) as a model system to study ferromagnetism in moir\'e flatbands.  
The continuum moir\'e Hamiltonian~\cite{Bistritzer2011} of TBG  is given by
\begin{equation}
\mathcal{H}_{\tau}=\begin{pmatrix}
h_{\tau b}(\kk) & T_{\tau }(\rr) \\
T^{\dagger}_{\tau}(\rr) & h_{\tau t}(\kk)
\end{pmatrix},
\label{Hmoire}
\end{equation}
where $\rr$ and $\kk$ are respectively position and momentum operators, and  $\tau=\pm$ is the valley index for $\pm K$ valleys that are related by time-reversal symmetry. In Eq.~\eqref{Hmoire}, the spin index is implicit because of spin SU(2) symmetry, and $T_{\tau }(\rr)$ is the periodic interlayer tunneling term \cite{Bistritzer2011}.
$h_{\tau b}$ and $h_{\tau t}$ are the Hamiltonians of the bottom ($\ell = b$) and top ($\ell = t$) layers:
\begin{equation}
h_{\tau \ell}(\kk) = h_{\tau \ell}^{(0)}(\kk) + \Delta_{\ell} \sigma_z/2,
\label{hDirac}
\end{equation}
where $h_{\tau \ell}^{(0)}(\kk)$ is the Dirac Hamiltonian for valley $\tau$ and layer $\ell$, and the additional term $\Delta_{\ell} \sigma_z/2$ is the sublattice potential difference in layer $\ell$. Here $\sigma_z$ is one of the Pauli matrices in the sublattice space. The potentials $\Delta_{b}$ ($\Delta_{t}$) are generated when TBG is in close alignment to the bottom (top) hexagonal boron nitride layers \cite{sharpe2019emergent,serlin2020intrinsic}, break the $\hat{C}_{2z}$ symmetry, and induce finite Berry curvatures for moir\'e bands. We take $\Delta_{b, t}$ as phenomenological parameters, while the values of other parameters in $\mathcal{H}_{\tau}$ are given in Ref.~\onlinecite{Wu2020Collective}.

\begin{figure}[t]
	\includegraphics[width=1\columnwidth]{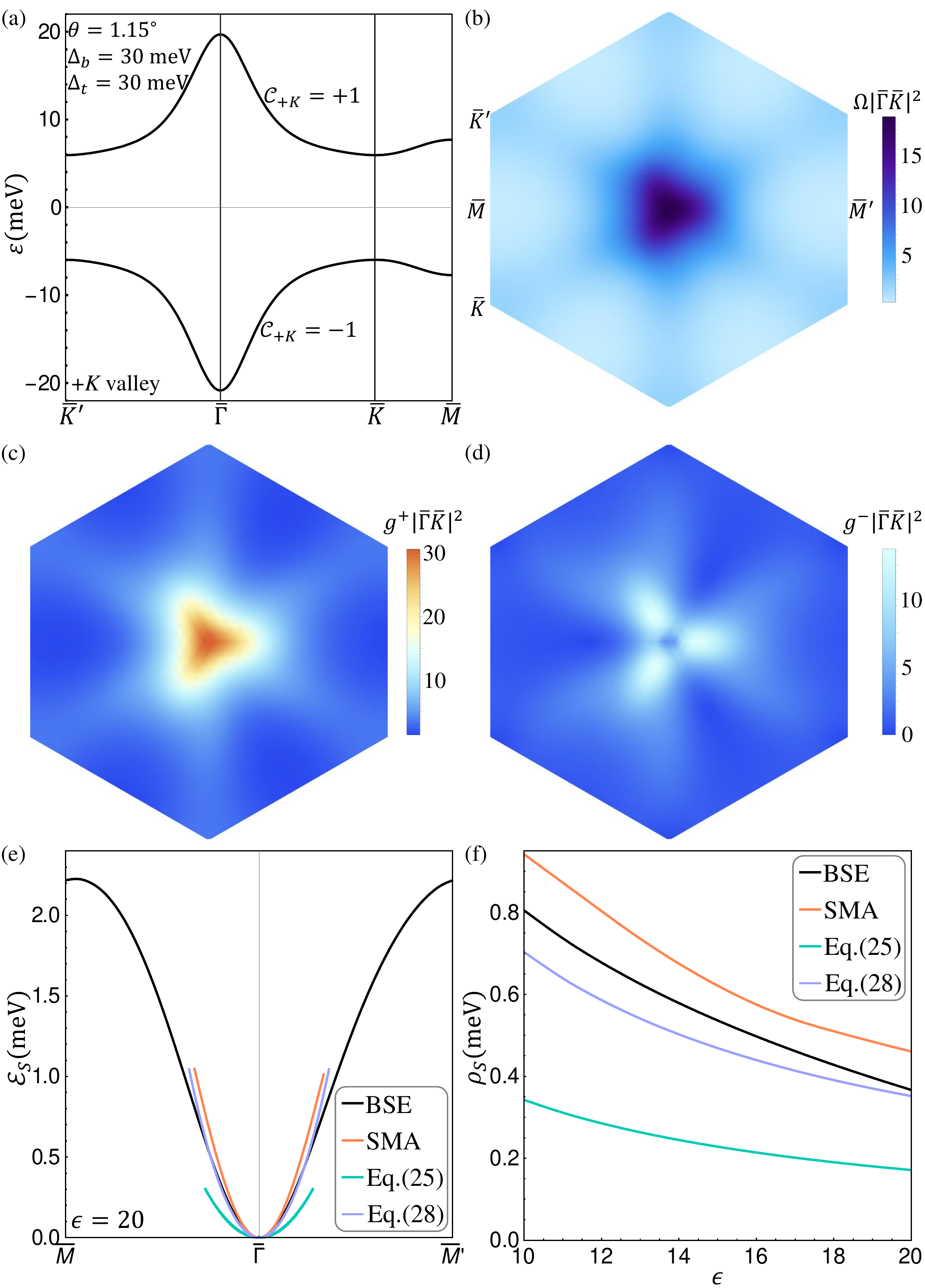}
	\caption{(a) Moir\'e band structure for $+K$ valley states in TBG with $\theta=1.15^{\circ}$ and $\Delta_{b}=\Delta_{t}=30$meV. The first conduction and valence band in $+K$ valley have Chern numbers of $+1$ and $-1$, respectively. (b) $\Omega_{\kk}$, (c) $g_{\kk}^+$ and (d) $g_{\kk}^-$ of the first conduction band in (a). $\Omega_{\kk}$ is the Berry curvature, and $g_{\kk}^{\pm}$ characterize the quantum metric. The plots are within the moir\'e Brillouin zone. (e) The spin wave dispersion and (f) the spin stiffness for the ferromagnetic state at the filling factor $\nu=3$. In (e) and (f), different curves are obtained using different approaches, with BSE and SMA denoting respectively the Bethe-Salpeter equation of Eq.~\eqref{SpinWave} and the single mode approximation of Eq.~\eqref{SWEn}.}
	\label{Fig:band_double}
\end{figure}

\begin{figure}[t]
	\includegraphics[width=1\columnwidth]{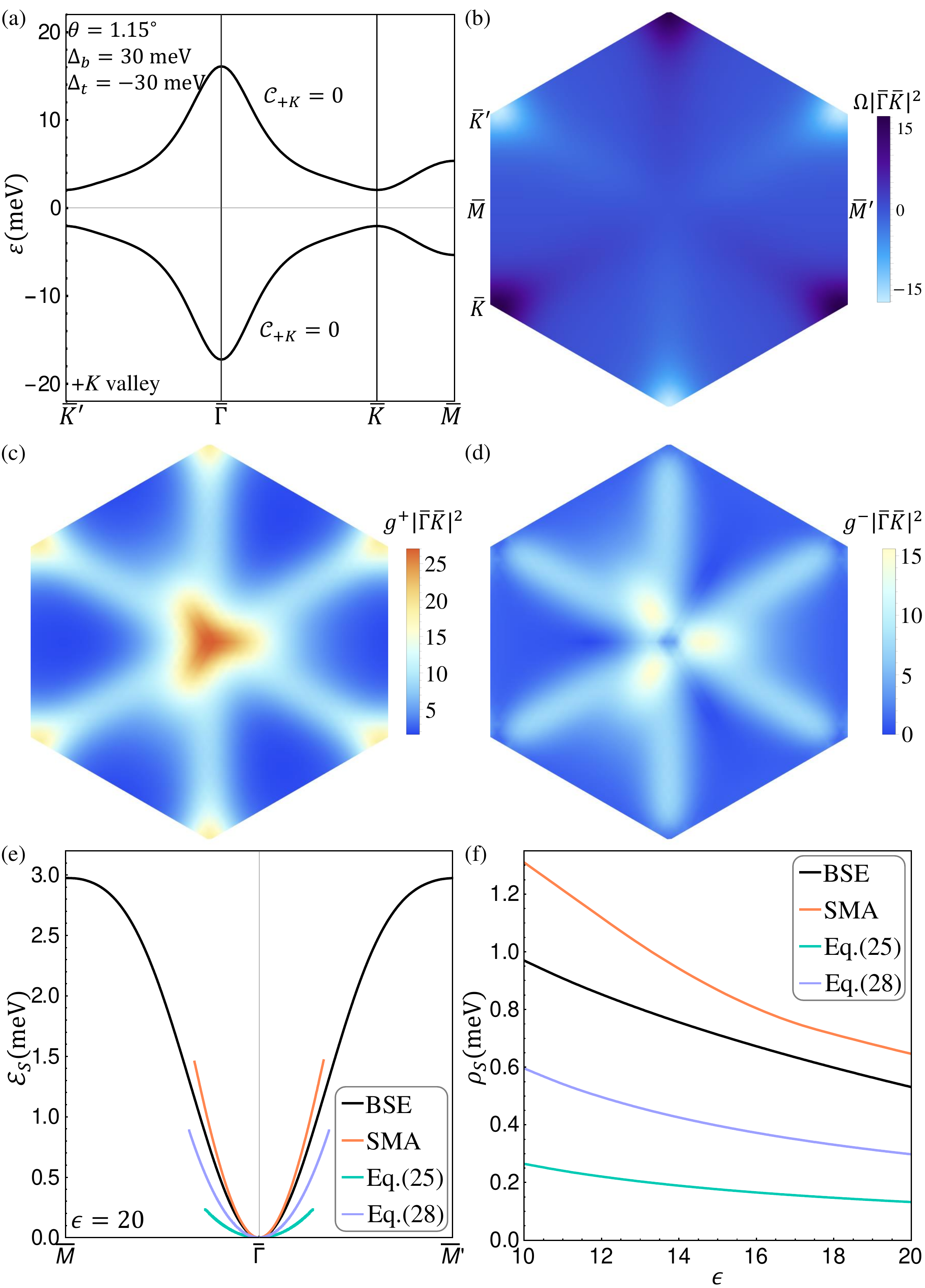}
	\caption{Similar plots as Fig.~\ref{Fig:band_double}. Important differences are that $\Delta_{b}=-\Delta_{t}$, and that Chern numbers for the first conduction and valence band are 0.}
	\label{Fig:band_opposite}
\end{figure}

Representative band structures of $\mathcal{H}_\tau$ are shown in Figs.~\ref{Fig:band_double}(a) and \ref{Fig:band_opposite}(a), respectively, for $\Delta_{b}=\Delta_{t}$ and $\Delta_{b}=-\Delta_{t}$. Because of the finite $\Delta_{b,t}$, Dirac cones located at $\bar{K}$ and $\bar{K}'$ (corners of the moir\'e Brillouin zone) are gapped out, and the first moir\'e conduction and valence bands are energetically separated. The Chern number $\mathcal{C}_{+K}$ of the first moir\'e conduction (valence) band in $+K$ valley is $+1$ ($-1$) for $\Delta_{b}=\Delta_{t}$, but $0$ ($0$) for $\Delta_{b}=-\Delta_{t}$, because of different patterns in the Berry curvatures, as shown in Figs.~\ref{Fig:band_double}(b) and \ref{Fig:band_opposite}(b). For $-K$ valley, the corresponding Chern number is  $\mathcal{C}_{-K}=-\mathcal{C}_{+K}$ following the time-reversal symmetry.

The low-energy moir\'e bands in TBG with a twist angle $\theta$ around $1^{\circ}$ have a narrow bandwidth ($\sim 10$ meV). The characteristic Coulomb interaction energy scale $E_\text{C}=e^2/(\epsilon a_M)$, with $\epsilon$ being the background dielectric constant and $a_M$ the moir\'e period, can be comparable to the bandwidth, and therefore, can give rise to strong correlation physics. We consider electron density at which the first moir\'e conduction bands are partially filled, and study an interacting model projected onto the first moir\'e conduction band states, which are separated from other bands when $\Delta_{b,t}$ are finite. This approximation of neglecting other bands is necessary for the later analytical study presented in Sections~\ref{sec:spinwave} and ~\ref{sec:skyrmion}.
The projected Hamiltonian $H$, including both the single-particle part $H_0$ and the interacting part $H_1$, is given by
\begin{align}
&H_0= \sum_{\kk, \tau, s} \varepsilon_{\kk,\tau} c^{\dagger}_{\kk,\tau, s} c_{\kk,\tau, s}\\
&H_1=\frac{1}{2A}\sum V_{\kk_1 \kk_2 \kk_3 \kk_4}^{(\tau \tau')}
c^{\dagger}_{\kk_1,\tau, s} c^{\dagger}_{\kk_2,\tau', s'} c_{\kk_3,\tau', s'} c_{\kk_4,\tau, s}, \\
&V_{\kk_1 \kk_2 \kk_3 \kk_4}^{(\tau \tau')} = \sum_{\qq} V(\qq) O_{\kk_1 \kk_4}^{(\tau)} (\qq) O_{\kk_2 \kk_3}^{(\tau')}(-\qq), \\
&O_{\kk \kk'}^{(\tau)} (\qq) =  \int d\rr e^{i \qq\cdot \rr} \Phi_{\kk,\tau}^{*}(\rr) \Phi_{\kk',\tau}(\rr),
\end{align}
where $c^{\dagger}_{\kk,\tau, s}$, $\varepsilon_{\kk,\tau}$ and $\Phi_{\kk,\tau}$ are respectively the electron creation operation, single-particle band energy, and wave function for the first conduction band states with valley index $\tau$, spin label $s$,  and momentum $\kk$. Here $\kk$ is measured relative to the moir\'e Brillouin zone center $\bar{\Gamma}$ point, and $\varepsilon_{\kk,\tau}=\varepsilon_{-\kk,-\tau}$ and $\Phi_{\kk,\tau}$=$\Phi^*_{-\kk,-\tau}$ because of time-reversal symmetry . In $H_1$, $A$ is the system area, $O_{\kk \kk'}^{(\tau)}(\qq)$ is the plane-wave matrix element, and $V(\qq)$ is the screened Coulomb potential $2\pi e^2 \tanh(q d)/(\epsilon q)$,
where $\epsilon$ is the effective dielectric constant, and $d$ is the vertical distance between TBG and top(bottom) metallic gates. In numerical calculations, we take $d$ to be 40 nm, which is a typical experimental value\cite{serlin2020intrinsic}.

\begin{figure}[t]
	\includegraphics[width=0.8\columnwidth]{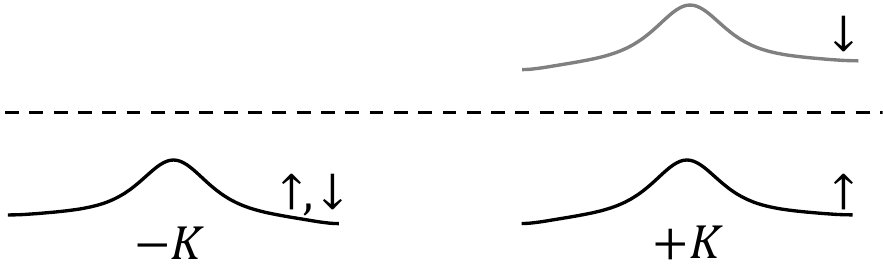}
	\caption{Schematic illustration of the $\nu=3$ ferromagnetic insulator with full spin and valley polarization.}
	\label{Fig:ferromagnet}
\end{figure}

The interaction matrix $V_{\kk_1 \kk_2 \kk_3 \kk_4}^{(\tau \tau')}$ is finite only  when $\kk_1+\kk_2-\kk_3-\kk_4$ is equal to $\boldsymbol{b}$, where $\boldsymbol{b}$ can be any moir\'e reciprocal lattice vectors (including the $\boldsymbol{0}$ vector), because of moir\'e translation symmetry. Similarly, $\qq$ in the matrix element $O_{\kk \kk'}^{(\tau)}(\qq)$ can differ from $\kk-\kk'$ by the vector $\bb$. 

The Hamiltonian $H$ respects threefold rotational symmetry, spin SU(2), valley U(1) symmetry and spinless time reversal symmetry. Furthermore, $H$ is invariant under the following gauge transformation: 
\begin{equation}
\begin{aligned}
\Phi_{\kk,\tau}(\rr) & \rightarrow e^{i\tau \varphi_{\kk}} \Phi_{\kk,\tau}(\rr), \\
c_{\kk,\tau,s}^{\dagger}& \rightarrow e^{i\tau \varphi_{\kk}} c_{\kk,\tau,s}^{\dagger} \\
c_{\kk,\tau,s}& \rightarrow e^{-i\tau \varphi_{\kk}} c_{\kk,\tau,s}
\end{aligned}
\label{eq:gauge}
\end{equation}
which reflects the fact that physical properties of the system should be independent of phase choices of single-particle wave functions. In Eq.~\eqref{eq:gauge}, $\varphi_{\kk}$ represents an arbitrary phase that can depend on $\kk$ but is independent of position $\rr$.

Motivated by the experimental observation of ferromagnetic insulators in TBG and related moir\'e systems, we use Hartree-Fock (HF) approximation to find the mean-field solutions of the Hamiltonian $H$. In the HF decomposition, we allow spin and valley polarization, which leads to the following mean-field Hamiltonian
\begin{equation}
\begin{aligned}
H_{\text{MF}} =& \sum_{\kk, \tau, s} E_{\kk,\tau, s} c^{\dagger}_{\kk,\tau, s} c_{\kk,\tau, s},\\
E_{\kk,\tau, s} =& \varepsilon_{\kk,\tau}+\frac{1}{A} \sum_{\kk', \tau', s'} V_{\kk \kk' \kk' \kk}^{(\tau \tau')} n_F(E_{\kk',\tau', s'})\\
&-\frac{1}{A} \sum_{\kk'} V_{\kk \kk' \kk \kk'}^{(\tau \tau)} n_F(E_{\kk',\tau, s}),
\label{HFM}
\end{aligned}
\end{equation}
where the quasiparticle energy $E_{\kk,\tau, s}$ includes the moir\'e band  energy $\varepsilon_{\kk,\tau}$ as well as the HF self energies, and $n_F$ represents the Fermi-Dirac occupation number.

\begin{figure}[t]
	\includegraphics[width=1\columnwidth]{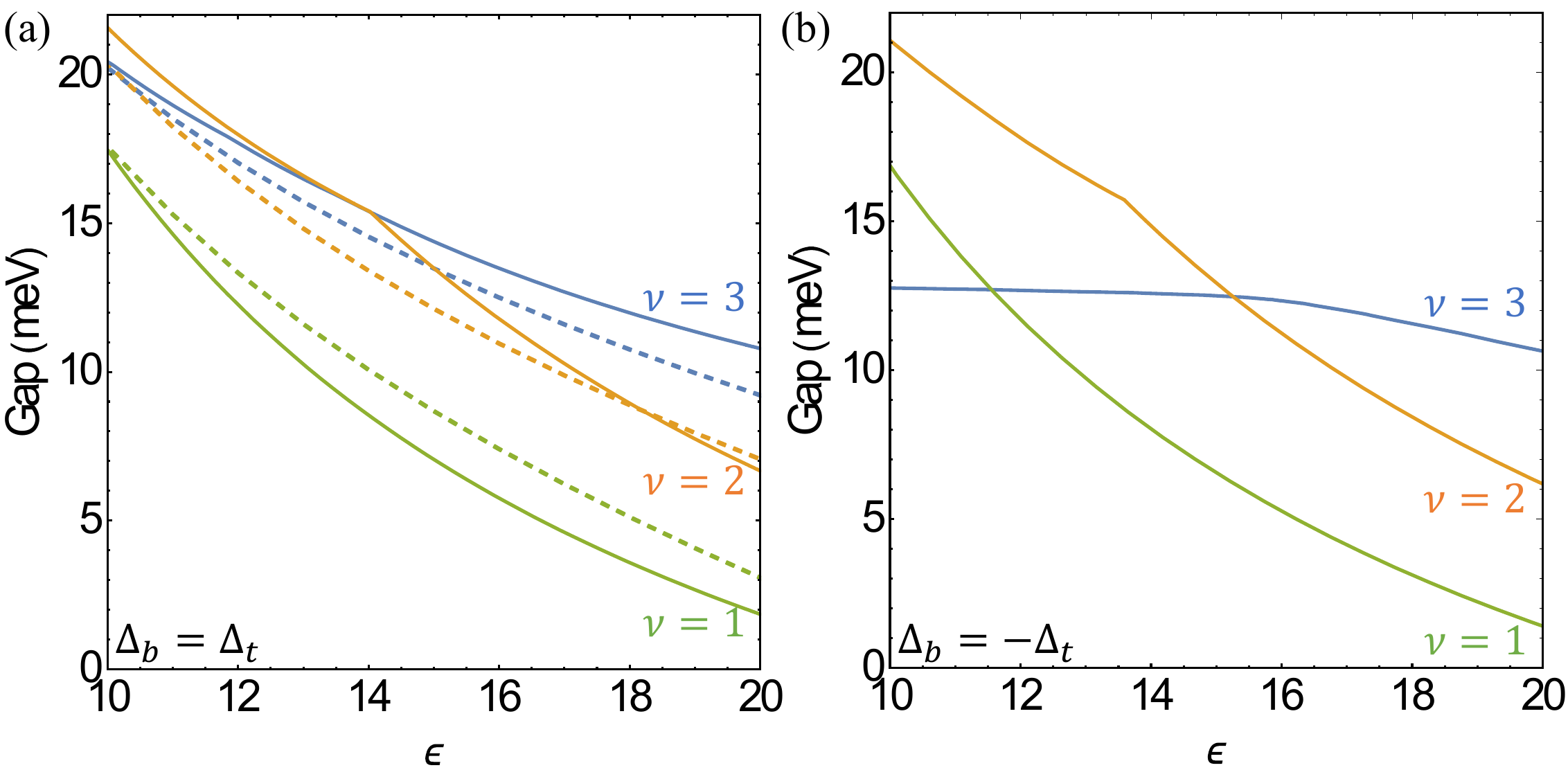}
	\caption{Charged excitation gap as a function of dielectric constant $\epsilon$ for different integer filling factors $\nu=1$, 2 and 3. Parameter values in (a) and (b) are, respectively, the same as those used in Figs.~\ref{Fig:band_double} and \ref{Fig:band_opposite}. The solid lines represent the HF gap $\Delta_{\text{HF}}$. In (a), the dashed lines plot $\Delta_{\text{pair}}$, which is the skyrmion-antiskyrmion pair energy. $\Delta_{\text{pair}}$ is estimated by using the spin stiffness extracted from the Bethe-Salpeter equation.}
	\label{Fig:HFGap}
\end{figure}

We define the electron filling factor $\nu$ as $n/n_0$, where $n$ is the electron density and $n_0$ the density for one electron per moir\'e unit cell. A full filling of the first moir\'e conduction bands corresponds to $\nu=4$, taking into account the spin and valley degeneracies. We consider commensurate filling factors $\nu=1$, 2 and 3, and make the ansatz that $\nu$ bands from the fourfold  band manifold are occupied and the remaining $4-\nu$ bands are unoccupied, which give rise to interaction driven spin and/or valley polarized insulators (zero temperature is assumed in this work). Here spin polarized states {\it spontaneously} break the spin SU(2) symmetry, but valley polarized states do {\it not} break the valley U(1) symmetry since the number of electrons associated with each valley remains conserved in the ansatz. The HF quasiparticle energy $E_{\kk,\tau, s}$ is calculated based on the above ansatz. We mention that this HF theory at the commensurate filling factors can be exact when all other bands can be ignored (i.e. if all the other bands are well-separated in energy), which is similar to the corresponding exactness of quantum Hall ferromagnetism in the single Landau level limit \cite{Moon1995}. The interaction-induced HF gap $\Delta_{\text{HF}}$ that separates occupied and unoccupied bands is shown in Fig.~\ref{Fig:HFGap} for the three integer filling factors  $\nu=1$, 2 and 3. As $\Delta_{\text{HF}}$ is finite for the parameter space that we explore, the spin and/or valley polarized states at  $\nu=1$, 2 and 3 can indeed be insulating.

\section{Spin wave}
\label{sec:spinwave}

We focus on the spin polarized insulators and their collective excitation spectrum. The spin magnon spectrum hosts a gapless spin wave mode, which is the Goldstone mode associated with the spontaneously broken SU(2) symmetry. In the following, we present different approaches to calculate the spin wave energy, using, respectively, the Bethe-Salpeter equation, the single-mode approximation, and an analytical theory. 

\subsection{Bethe-Salpeter Equation}
For definiteness, we consider the spin and valley maximally polarized state at $\nu=3$, and use $|\uparrow\rangle$ to denote the state in which only the valley $+K$ and  spin $\downarrow$  band is unoccupied, as illustrated in Fig.~\ref{Fig:ferromagnet}. Here $|\uparrow\rangle$ is an {\it exact} eigenstate of the many-body Hamiltonian $H$. The magnon state with intravalley spin flip on top of the $|\uparrow\rangle$ state can be parameterized as follows
\begin{equation}
|\QQ\rangle_S = \sum_{\kk} z_{\kk,\QQ} c_{\kk+\QQ,+,\downarrow}^{\dagger} c_{\kk,+,\uparrow} |\uparrow\rangle
\label{magnon}
\end{equation}
where $\QQ$ is the center-of-mass momentum of the magnon. Variation of the magnon energy with respect to the parameters $z_{\kk,\QQ}$ leads to the following eigenvalue equation
\begin{equation}
\begin{aligned}
&\mathcal{E}_S(\QQ)z_{\kk,\QQ}=\sum_{\kk'} \mathcal{H}_{\kk \kk'}^{(\QQ)}z_{\kk',\QQ},\\
&\mathcal{H}_{\kk \kk'}^{(\QQ)}= (E_{\kk+\QQ,+,\downarrow}-E_{\kk,+,\uparrow})\delta_{\kk,\kk'}-\frac{1}{A}V^{(++)}_{\kk' (\kk+\QQ) (\kk'+\QQ) \kk},
\end{aligned}
\label{SpinWave}
\end{equation}
where the matrix $\mathcal{H}_{\kk \kk'}^{(\QQ)}$ includes the quasiparticle energy cost of creating an electron-hole pair and the attractive interactions between electrons and holes. Equation~(\ref{SpinWave}) represents the Bethe-Salpeter equation for spin magnons in our system, and $\mathcal{E}_S(\QQ)$ is the magnon energy.

The matrix $\mathcal{H}_{\kk \kk'}^{(\QQ)}$ is not invariant (except at $\QQ = \boldsymbol{0}$)  under the gauge transformation in Eq.~\eqref{eq:gauge}. However, the characteristic polynomial of $\mathcal{H}_{\kk \kk'}^{(\QQ)}$  only involves product of wave function overlaps along closed loops in the momentum space, making the eigenvalues (i.e.,the magnon energy) gauge invariant. The eigenvectors of $\mathcal{H}_{\kk \kk'}^{(\QQ)}$ is gauge dependent, and $z_{\kk,\QQ}$ transforms to $z_{\kk,\QQ} \exp[i (\varphi_{\kk}-\varphi_{\kk+\QQ})]$ following Eq.~\eqref{eq:gauge}.

We calculate the magnon energy by numerically diagonalizing the matrix $\mathcal{H}_{\kk \kk'}^{(\QQ)}$, and show the magnon spectrum in Figs.~\ref{Fig:band_double}(e) and \ref{Fig:band_opposite}(e). In Fig.~\ref{Fig:band_double}, the non-interacting conduction bands have a finite {\it valley} Chern number, and the interaction driven ferromagnetic state $|\uparrow \rangle$ (i.e., the spin and valley polarized state at $\nu=3$) carries a {\it net} Chern number of $-1$, which leads to the quantum anomalous Hall effect. By contrast, the conduction bands in Fig.~\ref{Fig:band_opposite} are topologically trivial, and the corresponding ferromagnetic state $|\uparrow \rangle$ is also topologically trivial. Despite of the distinct topological characters, the magnon spectrum for these two cases look very similar: (1) the magnon energy shown in Figs.~\ref{Fig:band_double} and \ref{Fig:band_opposite} is nonnegative, which indicates the stability of the ferromagnetic state $|\uparrow \rangle$ against spin magnon excitations; (2) the lowest-energy spin magnon mode, i.e., the spin wave mode,  is gapless at $\QQ=\boldsymbol{0}$, as required by Goldstone's theorem.

The gapless Goldstone mode at $\QQ=\boldsymbol{0}$ can be constructed exactly.  The Hamiltonian $H$ respects the spin SU(2) symmetry, and therefore, commutes with the spin lowering operator
\begin{equation}
S^{-} = \sum_{\kk, \tau }  c_{\kk,\tau,\downarrow}^{\dagger} c_{\kk,\tau,\uparrow}.
\label{spinminus}
\end{equation} 
Applying $S^{-}$ to the $|\uparrow\rangle$ generates an eigenstate of $H$,
\begin{equation}
|\QQ=\boldsymbol{0}\rangle_\SW =  S^{-} | \uparrow \rangle = \sum_{\kk}   c_{\kk,+,\downarrow}^{\dagger} c_{\kk,+,\uparrow} |\uparrow \rangle,
\label{SWQ0}
\end{equation}
which is degenerate with the $|\uparrow\rangle$ state and represents the gapless mode in the spin magnon spectrum. Equation~\eqref{SWQ0} indicates that $z_{\kk, \QQ}=1$ for all $\kk$ is an exact zero-energy solution of $\mathcal{H}_{\kk \kk'}^{(\QQ)}$ at $\QQ=\boldsymbol{0}$, which can be confirmed explicitly.

\subsection{Single Mode Approximation}
We present a single mode approximation for the spin wave  by generalizing the spin lowering operator in Eq.~\eqref{spinminus} from the zero momentum to a finite momentum $\QQ$ 
\begin{equation}
\begin{aligned}
S_{\QQ}^{-} & = \sum_{\kk, \tau } M_{\kk+\QQ, \kk}^{(\tau)} c_{\kk+\QQ,\tau,\downarrow}^{\dagger} c_{\kk,\tau,\uparrow},\\
M_{\kk+\QQ, \kk}^{(\tau)}  & = \int d\rr e^{i \QQ \cdot \rr} \Phi_{\kk+\QQ,\tau}^{*}(\rr) \Phi_{\kk,\tau}(\rr) \\
& = \langle u_{\kk+\QQ, \tau} | u_{\kk,\tau}\rangle,
\end{aligned}
\end{equation}
where the plane-wave matrix element $M_{\kk+\QQ, \kk}^{(\tau)}$ makes the operator $S_{\QQ}^{-}$ gauge invariant, and $| u_{\kk,\tau}\rangle$ is  the periodic part of the wave function defined as $\exp(-i \kk \cdot \rr) \Phi_{\kk,\tau}(\rr)$. Applying the $S_{\QQ}^{-}$ operator to $|\uparrow\rangle$ generates the approximate spin wave mode at momentum $\QQ$
\begin{equation}
\begin{aligned}
|\QQ\rangle_\SW & = S_{\QQ}^{-} | \uparrow \rangle \\
&= \sum_{\kk}  M_{\kk+\QQ, \kk} c_{\kk+\QQ,+,\downarrow}^{\dagger} c_{\kk,+,\uparrow} |\uparrow \rangle.
\end{aligned}
\label{QQSW}
\end{equation}
where $M_{\kk+\QQ, \kk} \equiv M_{\kk+\QQ, \kk}^{(+)} $.  The ansatz in Eq.~\eqref{QQSW} is to replace the variation parameter $z_{\kk,\QQ}$ by $M_{\kk+\QQ, \kk}$. The energy of $|\QQ\rangle_\SW$ measured relative to that of the $|\uparrow\rangle$ state gives the approximate spin wave energy:
\begin{equation}
\mathcal{E}_{\SW}(\QQ) \approx \frac{\sum_{\kk, \kk'} M_{\kk+\QQ, \kk}^* \mathcal{H}_{\kk \kk'}^{(\QQ)} M_{\kk'+\QQ, \kk'}  }{\sum_{\kk} |M_{\kk+\QQ, \kk}|^2}.
\label{SWEn}
\end{equation}
At small $\QQ$, the spin wave energy calculated using Eq.~\eqref{SWEn} is in semiquantitative agreement with that obtained from the Bethe-Salpeter equation [Eq.~\eqref{SpinWave}], as shown in Figs.~\ref{Fig:band_double}(e) and \ref{Fig:band_opposite}(e). This agreement indicates that Eq.~\eqref{QQSW} represents a good ansatz for the spin wave mode.

\subsection{Effects of quantum geometry on spin stiffness}
In order to derive an analytical expression for the spin stiffness, we keep terms in  $\mathcal{E}_{\SW}(\QQ)$ that contribute up to second order of $\QQ$ as follows
\begin{equation}
\begin{aligned}
\mathcal{E}_{\SW}(\QQ) \approx \frac{1}{n_0 A^2} \sum_{\kk, \pp} \Big[ V_{\kk, \kk+\pp, \kk, \kk+\pp}^{(+ +)} |M_{\kk+\QQ,\kk}|^2
\\-V_{\kk, \kk+\pp+\QQ, \kk+\QQ, \kk+\pp}^{(+ +)} M_{\kk+\pp+\QQ, \kk+\pp}^*  M_{\kk+\QQ,\kk}
 \Big],
\end{aligned}
\label{ESWQQ2}
\end{equation}
where $n_0$, as defined above, is the density for one electron per moir\'e unit cell, and $n_0 A$ counts the total number of moir\'e unit cells in the system. To second order of $\QQ$, the spin wave energy $\mathcal{E}_{\SW}(\QQ)$ in Eq.~\eqref{ESWQQ2} is determined by the interaction potential $V(\qq)$ and the Bloch wave function $\Phi_{\kk,\tau}(\rr)$, but is independent of the single-particle band energy $\varepsilon_{\kk,\tau}$. In order to make further analytical progress, we neglect Umklapp scattering terms in the interaction matrix element,
\begin{equation}
\begin{aligned}
&V_{\kk, \kk+\pp+\QQ, \kk+\QQ, \kk+\pp}^{(+ +)}  \\
 = & \sum_{\bb} V(\pp+\bb) O_{\kk , \kk+\pp}^{(+)} (-\pp-\bb) O_{\kk+\pp+\QQ, \kk+\QQ}^{(+)}(\pp+\bb) \\
\approx &  V(\pp) M_{\kk, \kk+\pp} M_{\kk+\pp+\QQ, \kk+\QQ},
\end{aligned}
\label{VkkpQ}
\end{equation}
where $\bb$ represents moir\'e reciprocal lattice vectors, and the last line neglects terms with $\bb \neq \boldsymbol{0}$. The approximation in Eq.~\eqref{VkkpQ} could be justified by noting that the Coulomb potential $V(\qq)$ is sharply peaked around $\qq=\boldsymbol{0}$. Moreover, the wave function $\Phi_{\kk,\tau}(\rr)$ varies smoothly in real space (the characteristic length scale is the moir\'e period $a_M$), and therefore, the matrix element  $O_{\kk+\pp , \kk}^{(+)} (\pp+\bb)$, which is the Fourier transform of $\Phi_{\kk+\pp,+}^* (\rr) \Phi_{\kk,+} (\rr)$, should be a decreasing function of the momentum transfer $\pp+\bb$.   

\begin{figure}[t]
	\includegraphics[width=1\columnwidth]{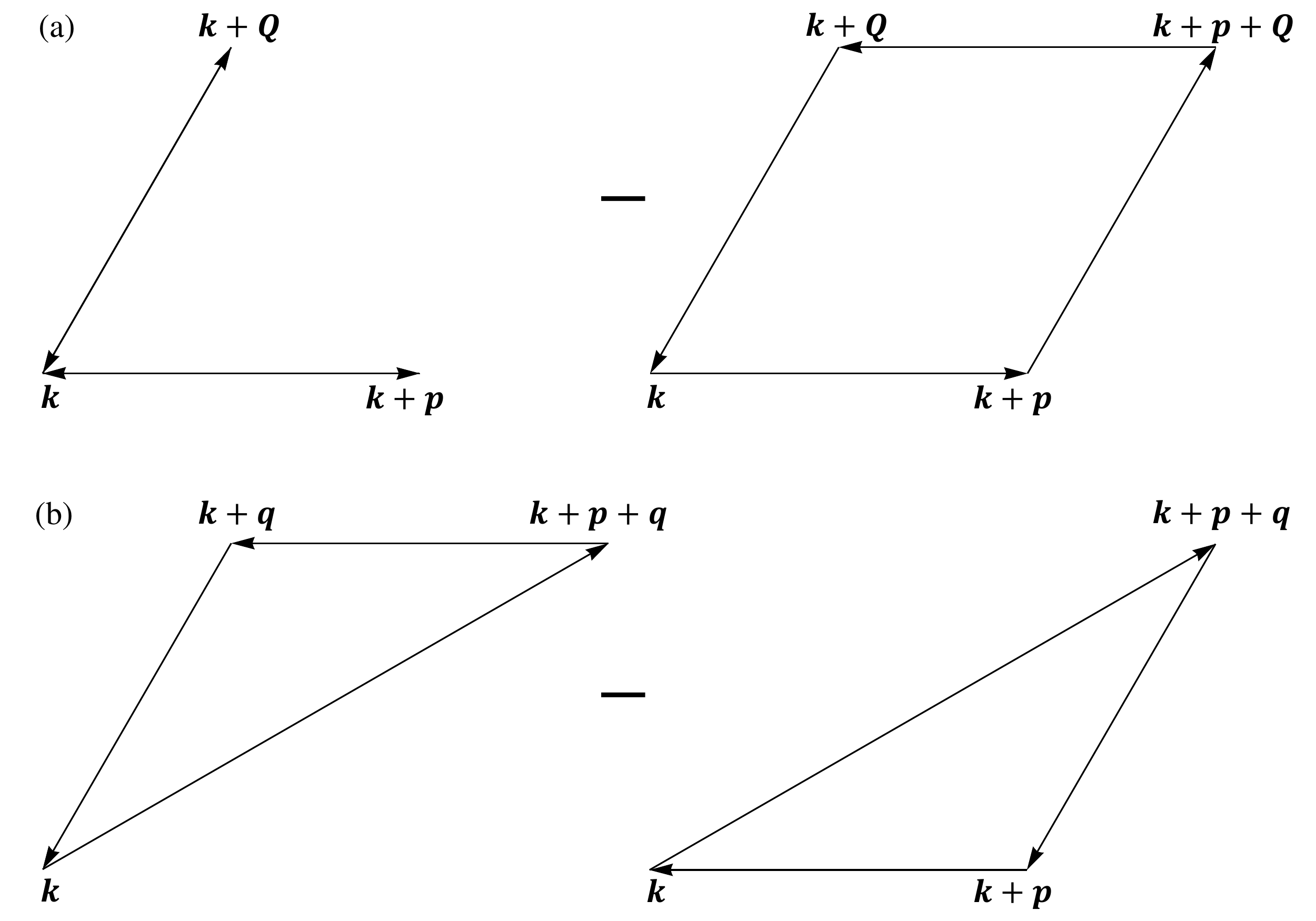}
	\caption{ Momentum loops that appear in (a) the spin wave energy of Eq.~\eqref{EWSQW} and (b) the spin-charge relation of Eq.~\eqref{spincharge}. Different loops enclose different Berry curvatures, which is the essential reason why Berry curvatures play a vital role in determining the spin wave energy and the spin-charge relation.}
	\label{Fig:WilsonLoop}
\end{figure}

With Eq.~\eqref{VkkpQ}, $\mathcal{E}_{\SW}(\QQ)$ can then be expressed as
\begin{equation}
\begin{aligned}
\mathcal{E}_{\SW}(\QQ) \approx \frac{1}{n_0 A^2} \sum_{\kk, \pp}  V(\pp) \Big[ \mathcal{W}(\kk,\kk+\pp,\kk,\kk+\QQ,\kk) \\ - \mathcal{W}(\kk,\kk+\pp,\kk+\pp+\QQ,\kk+\QQ,\kk)
\Big].
\end{aligned}
\label{EWSQW}
\end{equation}
where $\mathcal{W}$, representing the product of $M_{\kk_1, \kk_2} $ along a closed path in the momentum space. Momentum loops that appear in Eq.~\eqref{EWSQW} are illustrated in Fig.~\ref{Fig:WilsonLoop}(a). To be explicit, the definition of $\mathcal{W}$ is
\begin{equation}
\mathcal{W}(\kk_1,\kk_2,...,\kk_N,\kk_{N+1}=\kk_1) = \prod_{j=1}^{N} M_{\kk_j, \kk_{j+1}}~.
\label{Wilsonloopdef}
\end{equation}

The gauge-invariant quantity $\mathcal{W}$  can be expressed in terms of geometric quantities of the Bloch band
\begin{equation}
\begin{aligned}
&\mathcal{W}(\kk_1,\kk_2,...,\kk_N,\kk_{N+1}=\kk_1) & \\
\approx &  \exp [i \sum_{\kk_j} \mathcal{A}_{\alpha}(\kk_j) \delta k_\alpha-\frac{1}{2} \sum_{\kk_j} \hat{g}_{\alpha \beta}(\kk_j) \delta k_\alpha \delta k_\beta  ]\\
\approx & \exp [i \int_{\Gamma_L} \Omega_{\kk} d^2 \kk -\frac{1}{2} \sum_{\kk_j} \hat{g}_{\alpha \beta}(\kk_j) \delta k_\alpha \delta k_\beta],
\end{aligned}
\label{Wilsonloop}
\end{equation}
where  $\Gamma_L$ is the interior enclosed by the loop $L$ formed by $\kk_j$ for $j=1,...,N+1$, and $\delta \kk =\kk_{j+1}-\kk_j$.  Here $\mathcal{A}_{\alpha}(\kk)$ is the Berry connection, $\Omega_{\kk}$ is the Berry curvature and $\hat{g}(\kk)$ is the quantum metric tensor, with definitions respectively given by
\begin{equation}
\begin{aligned}
\mathcal{A}_{\alpha}(\kk) & = -i \langle u_{\kk} |\partial_{k_\alpha}| u_{\kk} \rangle, \\
\Omega_{\kk} & = \partial_{k_x} \mathcal{A}_{y}(\kk)-\partial_{k_y} \mathcal{A}_{x}(\kk),\\
\hat{g}_{\alpha \beta}(\kk) & = \text{Re} [\langle \partial_{k_\alpha} u_{\kk} | \partial_{k_\beta} u_{\kk} \rangle] - \mathcal{A}_{\alpha}(\kk)\mathcal{A}_{\beta}(\kk),
\end{aligned}
\end{equation}
where the valley index $\tau$ is understood to be $+$ and neglected for brevity. While the Berry connection $\mathcal{A}_{\alpha}$ is gauge dependent, the Berry curvature $\Omega$ and quantum metric tensor $\hat{g}$ are gauge invariant and characterize the quantum geometry of the Bloch bands. By definition, the tensor $\hat{g}$ is real and symmetric. Equation~\eqref{Wilsonloop} is derived by using the following expansion 
\begin{equation}
\begin{aligned}
&\langle u_{\kk-\pp/2} | u_{\kk+\pp/2} \rangle \\
\approx & 1+ i p_\alpha \mathcal{A}_{\alpha}(\kk)- \frac{1}{2} p_\alpha p_\beta \text{Re} [\langle \partial_{k_\alpha} u_{\kk} | \partial_{k_\beta} u_{\kk} \rangle] \\
\approx & \exp[i p_\alpha \mathcal{A}_{\alpha}(\kk)-\frac{1}{2} \hat{g}_{\alpha \beta}(\kk)  p_\alpha p_\beta],
\end{aligned} 
\label{Infoverlap}
\end{equation}
where each approximation is valid up to second order of $\pp$. Equation~\eqref{Infoverlap} indicates that $\hat{g}$ acts as the metric that measures the ``quantum distance'', i.e., $1-|\langle u_{\kk-\pp/2} | u_{\kk+\pp/2} \rangle|^2$, between the two Bloch states. Therefore, the tensor $\hat{g}$ is dubbed as quantum metric, which is always semipositive definite. 
The exponential form in Eq.~\eqref{Infoverlap} is derived by assuming that $\pp$ is small. As argued in the above, the wave function $\Phi_{\kk,\tau}(\rr)$ varies smoothly in real space, and therefore, the overlap $\langle u_{\kk-\pp/2} | u_{\kk+\pp/2} \rangle$ should decay exponentially with $\pp$ also for large $\pp$. Because of this exponential suppression, the main contributions to the spin wave energy in Eq.~\eqref{EWSQW} come from terms with small momentum transfer $\pp$. With this justification, we adopt Eq.~\eqref{Infoverlap} beyond the small-$\pp$ regime when evaluating Eq~\eqref{EWSQW}.

By combining Eqs.~\eqref{EWSQW} and \eqref{Wilsonloop}, we obtain an analytical expression for the spin wave energy
\begin{equation}
\begin{aligned}
\mathcal{E}_{\SW}(\QQ) %\\
% \approx  & \frac{1}{n_0 A^2} \sum_{\kk, \pp}  \Big\{V(\pp) \exp \left(-\pp \hat{g}_{\kk} \pp^\text{T}-\QQ \hat{g}_{\kk} \QQ^\text{T} \right) \\
%&\,\,\,\,\,\,\,\,\,\,\,\,\,\,\,\,\,\,\,\,\,\,\,\,\,\,\times \Big[ 1-\exp[i \Omega_{\kk} (\pp \times \QQ)\cdot \hat{z}]
%\Big] \Big\} \\
\approx  \frac{1}{4 n_0 A^2} \sum_{\kk, \pp}  V(\pp) \exp(-\pp \hat{g}_{\kk} \pp^\text{T}) \Omega_{\kk}^2 \pp^2  \QQ^2
\end{aligned}
\label{ESWp2Q2}
\end{equation}
which is an expansion to second order in $\QQ$. 

The spin stiffness $\rho_s$ extracted from Eq.~\eqref{ESWp2Q2} by using the definition $\mathcal{E}_{\SW}(\QQ)=(2\rho_s/n_0)\QQ^2$ is
\begin{equation}
\rho_s \approx \frac{1}{8 A^2} \sum_{\kk, \pp}  \Omega_{\kk}^2 \pp^2 V(\pp) \exp(-\pp \hat{g}_{\kk} \pp^\text{T}),
\label{eq:rhosOVg}
\end{equation}
which shows that the absolute value of Berry curvature $\Omega_{\kk}$ contributes to  $\rho_s$, but the quantum metric $\hat{g}_{\kk}$ tends to suppress $\rho_s$. We emphasize that the sign of $\Omega_{\kk}$ plays no role in Eq.~\eqref{eq:rhosOVg}. This provides an explanation on why the spin stiffness is finite in both Figs.~\ref{Fig:band_double} and \ref{Fig:band_opposite}, where the Berry curvatures are finite for both cases but have drastically different sign structures in momentum space. 

The summation over $\pp$ in Eq.~\eqref{eq:rhosOVg} can be performed analytically by using the unscreened Coulomb potential $V(\pp)=2\pi e^2/(\epsilon |\pp|)$ and by extending the range of $\pp$ from the first moir\'e Brillouin zone to the full momentum space, which can be justified by noting the exponential decaying factor $\exp(-\pp \hat{g}_{\kk} \pp^\text{T})$. The resulting $\rho_s$ is
\begin{equation}
\rho_s \approx \frac{1}{4\sqrt{2\pi}} \frac{e^2}{\epsilon} \int \frac{d^2 \kk}{(2\pi)^2} \Omega_{\kk}^2 \frac{\mathscr{E} [2 g_{\kk}^{-}/( g^{+}_{\kk}+ g^{-}_{\kk})]}{(g^{+}_{\kk}- g^{-}_{\kk})\sqrt{g^{+}_{\kk} + g^{-}_{\kk}}},
\label{eq:rhosVOg}
\end{equation}
where $g_{\kk}^{\pm}= g_{\kk}^{(1)} \pm g_{\kk}^{(2)}$. Here $g_{\kk}^{(1)}$ and  $g_{\kk}^{(2)}$ are the two eigenvalues of the tensor $\hat{g}(\kk)$, with $g_{\kk}^{(1)} \geq g_{\kk}^{(2)} \geq 0$. The function $\mathscr{E}$ is the elliptic integral defined as 
\begin{equation}
\mathscr{E} (x) = \int_{0}^{\pi/2} d\phi \sqrt{1- x \sin^2 \phi } ~.
\end{equation}

We numerically calculate the quantum metric $\hat{g}$ based on Eq.~\eqref{Wilsonloop}, and show $g_{\kk}^{\pm}$ in Figs.~\ref{Fig:band_double} and \ref{Fig:band_opposite}. Similar to the Berry curvature $\Omega_{\kk}$, $g_{\kk}^{\pm}$ respects the symmetry in the momentum space. Particularly, $g_{\kk}^{-}$ vanishes at $\bar{\Gamma}$, $\bar{K}$ and $\bar{K}'$ points because of threefold rotation symmetry. Furthermore, there is an intrinsic lower bound, set by $|\Omega_{\kk}|$, on $g^{+}_{\kk}$, as described by the following inequality
\begin{equation}
g^{+}_{\kk} \equiv \text{Tr} \hat{g}_{\kk} \ge |\Omega_{\kk}|,
\label{eq:ineqgO}
\end{equation}
which has been proved in Ref.~\onlinecite{Roy2014Geometry}. We also provide a proof of Eq.~\eqref{eq:ineqgO} in Appendix~\ref{app}.

By approximating $g^{+}_{\kk}$ and $g^{-}_{\kk}$ to be, respectively,  $|\Omega_{\kk}|$ and 0, we can further simplify $\rho_s$ to be 
\begin{equation}
\rho_s \approx \frac{1}{8} \sqrt{\frac{\pi}{2}} \frac{e^2}{\epsilon} \int \frac{d^2 \kk}{(2\pi)^2} |\Omega_{\kk}|^{1/2} 
\label{eq:rhosVO}
\end{equation}
which depends only on the the Berry curvature $\Omega_{\kk}$ and the dielectric constant $\epsilon$. Equation~\eqref{eq:rhosVO} shows the direct analogy of moir\'e flatband ferromagnetism to Landau level quantum Hall ferromagnetism (see Appendix~\ref{app}) in the sense that the momentum integral in this equation can be interpreted as the effective inverse ``Landau radius'' for the moir\'e system although the effective magnetic field (associated with the ``Landau radius'' in the quantum Hall system) here is entirely a quantum geometric effect as there is no applied magnetic field in the moir\'e system.

Equations~\eqref{eq:rhosOVg}, \eqref{eq:rhosVOg} and \eqref{eq:rhosVO}, at different levels of approximation, express the spin stiffness $\rho_s$ analytically in terms of interaction potential and Bloch-band quantum geometry, which are the main results of this work. We show values of $\rho_s$ estimated using different approaches in Figs.~\ref{Fig:band_double}(f) and \ref{Fig:band_opposite}(f), and take $\rho_s$ calculated directly from the Bethe-Salpeter equation [Eq.~\eqref{SpinWave}] as the benchmark to check other approximations. The comparison can be summarized as follows. (1) The single mode approximation of Eq.~\eqref{SWEn} overestimates $\rho_s$, which is expected since the single mode state in Eq.~\eqref{QQSW} represents an approximate ansatz to the true spin wave state. (2) While both Eqs.~\eqref{eq:rhosVOg} and ~\eqref{eq:rhosVO} generally underestimates $\rho_s$, Eq.~\eqref{eq:rhosVO} can provide a better estimation despite the fact that it is a further approximation to Eq.~\eqref{eq:rhosVOg}. This is because $\exp(-\pp \hat{g}_{\kk} \pp^\text{T})$ overestimates the finite $\pp$ reduction of $|\langle u_{\kk-\pp/2} | u_{\kk+\pp/2} \rangle|^2$, and replacing $(g_{\kk}^+,g_{\kk}^-)$ by their lower bounds $(|\Omega_{\kk}|,0)$ partially cures this problem.

As shown by results in Figs.~\ref{Fig:band_double}(f) and \ref{Fig:band_opposite}(f), Eq.~\eqref{eq:rhosVO} provides a semiquantitative estimation of the spin stiffness, which indicates that Berry curvatures of Bloch bands can contribute significantly to stiffen the spin magnons. It is the absolute value of Berry curvatures, i.e., $|\Omega_{\kk}|$, that enter into the expression of $\rho_s$. Therefore, a topologically trivial band with a zero Chern number  can still support ferromagnetism, provided that the Berry curvatures are finite in momentum space as the case in Fig.~\ref{Fig:band_opposite}.

\section{Field theory and skyrmions}
\label{sec:skyrmion}
We present another approach to calculate the spin wave energy by constructing an effective field theory, which is based on the following spin texture state
\begin{equation}
\begin{aligned}
&| \mm(\rr)  \rangle  = \exp(- i F) | \uparrow \rangle, \\
& F = \int d \rr \left[ m_x(\rr) S_y(\rr)-m_y(\rr) S_x(\rr) \right]\\
&\,\,\,\,\, =\sum_{\qq} m_{\qq}^x S_{-\qq}^y-m_{\qq}^y S_{-\qq}^x,\\
&\bS(\rr) = \frac{1}{\sqrt{A}}\sum_{\qq} e^{-i\qq \cdot \rr} \bS_{\qq},\\
&\bS_{\qq} = \frac{1}{\sqrt{A}} \sum_{\kk,s_1, s_2} M_{\kk+\qq,\kk} \,\, c_{\kk+\qq,+,s_1}^{\dagger} \frac{\boldsymbol{\sigma}_{s_1 s_2}}{2} c_{\kk,+,s_2},
\end{aligned}
\end{equation}
where $\mm(\rr)$ represents a unit vector with small in-plane components $m_{x,y}$ and smooth spatial variations,  and $\bS(\rr)$ is the local spin operator projected to $+K$ valley. $\mm_{\qq}$ and $\bS_{\qq}$ are, respectively, the Fourier components of $\mm(\rr)$ and $\bS(\rr)$. The operator $\exp(- i F)$ rotates the local spin direction from $\hat{z}$ to $\mm(\rr)$. Thus, the state $| \mm(\rr)  \rangle$ has a slowly varying spin texture. By taking $m_{x,y}$ as small parameters, we can expand the energy of the spin texture state in powers of $F$
%\begin{equation}
%| \mm(\rr)  \rangle \approx (1- i F- \frac{1}{2} F^2) | \uparrow \rangle
%\end{equation}
\begin{equation}
\begin{aligned}
\mathcal{E}[\mm(\rr)] & = \langle \mm(\rr)| H | \mm(\rr)  \rangle - \langle \uparrow | H | \uparrow  \rangle \\
& \approx i \langle \uparrow | [F,H] | \uparrow  \rangle-\frac{1}{2}\langle \uparrow |[F, [F,H]] | \uparrow  \rangle,
\end{aligned}
\label{Energymm}
\end{equation}
where the first order term exactly vanishes. The second order term in  Eq.~\eqref{Energymm} gives rise to momentum space integral very similar to Eq.~\eqref{EWSQW}, and therefore, can be computed similarly. The resulting energy functional is given by
\begin{equation}
\begin{aligned}
\mathcal{E}[\mm(\rr)] &\approx \frac{\rho_s}{2} \sum_{\qq}  \qq^2 (m_{\qq}^{x}m_{-\qq}^{x}+m_{\qq}^{y}m_{-\qq}^{y}) \\
&= \frac{\rho_s}{2} \int d\rr \big\{[\nabla m_x(\rr)]^2+[\nabla m_y(\rr)]^2\big\},
\end{aligned}
\end{equation}
where $\rho_s$ is the spin stiffness with the same expression as Eq.~(\ref{eq:rhosOVg}).

The effective Lagrangian includes not only the energy functional $\mathcal{E}[\mm(\rr)]$ but also the kinetic Berry phase $\mathcal{B}_S$,
\begin{equation}
\begin{aligned} 
\mathcal{L}_S &= \mathcal{B}_S  - \mathcal{E}[\mm(\rr)]\\
\mathcal{B}_S &= \langle \mm(\rr)| i\hbar \partial_t | \mm(\rr)  \rangle \\
& \approx -\frac{\hbar n_0}{4} \int d\rr (m_x \partial_t m_y-m_y \partial_t m_x) 
\end{aligned}
\label{fieldtheory}
\end{equation}
where $t$ represents time. The field theory in Eq.~\eqref{fieldtheory} captures the low-energy and long-wavelength spin dynamics in the ferromagnet. The corresponding equation of motion has spin wave solutions, where magnetization $\mm(\rr)$    precesses around $\hat{z}$ direction with a wave vector $\QQ$ at frequency $\omega = (2\rho_s/n_0)\QQ^2/\hbar$. Here $\hbar \omega$ is exactly the spin wave energy. 

The Lagrangian $\mathcal{L}_S$ can be recast into spin rotation invariant form
\begin{equation}
\mathcal{L}_S=-\int d^2 \rr \Big\{\frac{\hbar n_0}{2} \boldsymbol{\mathcal{A}}[\mm]\cdot \partial_t \mm+\frac{\rho_s}{2} (\boldsymbol{\nabla} \mm)^2 \Big\},
\label{LS}
\end{equation}
which is the $O$(3) nonlinear sigma model. Here $\mathcal{A}[\mm]$ is the effective spin gauge field defined by $\boldsymbol{\nabla}_{\mm} \times \mathcal{A}[\mm] =\mm$. 

The $O$(3) nonlinear sigma model also supports another type of excitations, namely, skyrmions. We show that skyrmions carry an integer number of excess charge when the underlying Bloch band is topological with a nonzero Chern number. This physics is known in the quantum Hall regime \cite{SondhiSkyrmion,Moon1995}, and we generalize it from Landau levels to Bloch bands with nonuniform Berry curvatures. We first define a density operator as follows
\begin{equation}
\rho_{\qq} = \sum_{\kk,\tau,s} M_{\kk+\qq,\kk}^{(\tau)} c_{\kk+\qq,\tau,s}^{\dagger}  c_{\kk,\tau,s}.
\end{equation}
The excess charge in the spin texture state is then given by
\begin{equation}
\begin{aligned}
 \delta \rho_{\qq}  &\equiv \langle \mm(\rr)| \rho_{\qq} | \mm(\rr)  \rangle - \langle \uparrow | \rho_{\qq} | \uparrow  \rangle \\
 &= \frac{\mathcal{C}}{8 \pi A^{1/2}} \sum_{\pp} (\pp \wedge \qq)(\mm_{\pp+\qq} \wedge \mm_{-\pp}),
\end{aligned}
\label{drhoq}
\end{equation}
where $\pp \wedge \qq = (\pp \times \qq)\cdot \hat{z}$ and $\mathcal{C}=\int d\kk ~\Omega_{\kk}/(2\pi)$ is the Chern number. Equation~\eqref{drhoq} is derived by using the following spin-charge commutation
\begin{equation}
\begin{aligned}
&\langle \uparrow | [S_{-\pp-\qq}^{\alpha},[S_{\pp}^{\beta},\rho_{\qq}]]| \uparrow  \rangle \\
= & \frac{i \epsilon_{\alpha \beta}}{2A^{3/2}} \sum_{\kk} \big[
\mathcal{W}(\kk,\kk+\pp+\qq,\kk+\qq,\kk)\\
&\,\,\,\,\,\,\,\,\,\,\,\,\,\,\,\,\,\,\,\,\,\,\,\,\,\,
-\mathcal{W}(\kk,\kk+\pp+\qq,\kk+\pp,\kk)
\big]\\
\approx & - \frac{\epsilon_{\alpha \beta}}{2A^{3/2}} \left( \pp \wedge \qq \right) \sum_{\kk} \Omega_{\kk} \\
= & - \frac{\epsilon_{\alpha \beta}}{4\pi A^{1/2}} \left( \pp \wedge \qq \right) \mathcal{C},
\end{aligned}
\label{spincharge}
\end{equation}
where $\epsilon_{\alpha \beta}$ represents the antisymmetric tensor with $\epsilon_{xy}=-\epsilon_{yx}=1$, and $\mathcal{W}$ is defined in Eq.~\eqref{Wilsonloopdef} and evaluated using Eq.~\eqref{Wilsonloop}. A diagrammatic representation of Eq.~\eqref{spincharge} is shown in Fig.~\ref{Fig:WilsonLoop}(b).

The excess charge in real space is obtained by applying Fourier transformation to Eq.~\eqref{drhoq},
\begin{equation}
\begin{aligned}
 \delta \rho(\rr) &\equiv \frac{1}{\sqrt{A}} \sum_{\qq} e^{-i \qq \cdot \rr}  \delta \rho_{\qq}\\
& = -\frac{\mathcal{C}}{4\pi} \mm(\rr)\cdot [\partial_x \mm(\rr) \times \partial_y \mm(\rr)],
\end{aligned}
\end{equation}
which is the Chern number $\mathcal{C}$ times the Pontryagin index density (or topological charge density) of the spin texture. The total extra charge bound to a skyrmion is a quantized number determined by $\mathcal{C}$ and the skyrmion winding number $\mathcal{N}_w$, 
\begin{equation}
\begin{aligned}
\Delta N & \equiv \int d\rr  \delta \rho(\rr) = - \mathcal{C} \mathcal{N}_w,\\
\mathcal{N}_w & \equiv \frac{1}{4\pi}\int d\rr  \mm(\rr)\cdot [\partial_x \mm(\rr) \times \partial_y \mm(\rr)].
\end{aligned}
\end{equation}

Therefore, skyrmions are charged when the underlying Bloch bands carry nonzero Chern numbers. In the topological case, the charged excitaton gap for the ferromagnet is determined by the Hartree-Fock gap $\Delta_{\text{HF}}$ or the energy cost $\Delta_{\text{pair}}$ for creating a pair of skyrmions with opposite winding numbers $\mathcal{N}_w=\pm 1$, whichever is lower. From the $O$(3) nonlinear sigma model, we obtain $\Delta_{\text{pair}}=8\pi \rho_s$. As shown in Fig.~\ref{Fig:HFGap}, $\Delta_{\text{pair}}$ can be lower or higher in energy compared to $\Delta_{\text{HF}}$, depending on system details.

%\begin{align*}
%[\bS_{\qq},\rho_{\pp}]=\sum_{\kk} \big( & M_{\kk+\pp+\qq,\kk+\pp}M_{\kk+\pp,\kk} %\\
% &-M_{\kk+\pp+\qq,\kk+\qq}M_{\kk+\qq,\kk} \big) \\
% & c_{\kk+\pp+\qq,+,s_1}^{\dagger} \frac{\boldsymbol{\sigma}_{s_1 s_2}}{2} c_{\kk,+,s_2}
%\end{align*}

\section{Conclusion}
\label{sec:dis}

In summary, we present an analytical theory of spin stiffness, which elucidates the role of quantum geometry. 
We find that the spin stiffness $\rho_s$ is an increasing function of $|\Omega_{\kk}|$. An implication is that moir\'e bands with higher Chern numbers, as realized in twisted double bilayer graphene \cite{Wu2020Ferro} and also twisted monolayer-bilayer graphene \cite{polshyn2020nonvolatile,chen2020electrically}, could be more favorable for ferromagnetism.  
We note that $\rho_s$ can only characterize the spin wave dispersion in the long-wavelength limit ($\QQ \rightarrow \boldsymbol{0}$). The stability of ferromagnetism requires that the spin magnon spectrum is nonnegative in the full moir\'e Brillouin zone. Therefore, a positive spin stiffness is a necessary but not a sufficient criterion for the robustness of ferromagnetism. 

In addition to spin magnons, moir\'e flatband ferromagnetism can also have valley magnon excitations (i.e, intervalley excitons). Because valley polarized states do not break the valley U(1) symmetry, valley magnons are generically gapped, which is another criterion required for the stability of spin and valley polarized ferromagnets. This criterion is numerically verified for the ferromagnets studied in Figs.~\ref{Fig:band_double} and \ref{Fig:band_opposite}, by solving the Bethe-Salpeter equation for the valley magnons \cite{Wu2020Collective}. Whether the valley magnon energy can be analytically expressed in terms of band geometric quantities is an interesting open question that we leave for future study. 

Our work also brings out the natural deep connection between quantum Hall ferromagnetism and moir\'e flatband ferromagnetism, showing that the spin stiffness in the two cases have formally similar expressions.  In addition, the ferromagnetism in both cases becomes an exact solution within the HF theory as long as other bands in the moir\'e system (other Landau levels in the quantum Hall system) can be neglected.  Providing the direct connection of quantum geometry to the ferromagnetism in moir\'e systems is our important theoretical finding.

\section{acknowledgments}
F.W. thanks Y. Alavirad for stimulating discussions. This work is supported by the Laboratory for Physical Sciences.

\appendix
\section{QUANTUM GEOMETRY}
\label{app}

We present a proof of Eq.~\eqref{eq:ineqgO} in this Appendix. The Berry curvature and the quantum metric can be combined to define a quantum geometric tensor $\hat{\Lambda}$ as follows
\begin{equation}
\hat{\Lambda}_{\alpha \beta}(\kk) = \hat{g}_{\alpha \beta}(\kk)+\frac{i}{2} \epsilon_{\alpha \beta} \Omega_{\kk},
\end{equation}
where $\epsilon_{\alpha \beta}$ is the antisymmetric tensor. Here $\hat{\Lambda}$ is a hermitian matrix that can be organized into the following form
\begin{equation}
\begin{aligned}
\hat{\Lambda}_{\alpha \beta}(\kk)&=\langle \partial_{k_\alpha} u_{\kk} | \partial_{k_\beta} u_{\kk} \rangle -\mathcal{A}_{\alpha}(\kk)\mathcal{A}_{\beta}(\kk)\\
&=\langle \partial_{k_\alpha} u_{\kk} |(\hat{I}-\hat{P}_{\kk})| \partial_{k_\beta} u_{\kk} \rangle,
\end{aligned}
\end{equation} 
where $\hat{I}$ is the identity matrix and $\hat{P}_{\kk}$ is the projector $|u_{\kk} \rangle \langle u_{\kk}|$. $\hat{I}-\hat{P}_{\kk}$ represents another projector that is complementary to $\hat{P}_{\kk}$. The tensor $\hat{\Lambda}$ is the projection of $\hat{I}-\hat{P}_{\kk}$ onto the subspace spanned by $\{|\partial_{k_x} u_{\kk} \rangle, |\partial_{k_y} u_{\kk} \rangle \}$. The projector $\hat{I}-\hat{P}_{\kk}$ is semipositive definite, so is the tensor $\hat{\Lambda}$. Therefore,  $\text{Tr} \hat{\Lambda} \geq 0$ and  $\det \hat{\Lambda} \geq 0$ . Noting that $\text{Tr} \hat{\Lambda}=\text{Tr} \hat{g}$ and $\det \hat{\Lambda} =  - \Omega^2/4 +\det \hat{g}$, we obtain the following inequalities  
\begin{equation}
\text{Tr} \hat{g}_{\kk} \geq 0, \,\,\, 
\det \hat{g}_{\kk} \geq \frac{\Omega_{\kk}^2}{4}.
\end{equation}
Because $\hat{g}_{\kk} $ is a $2\times 2$ matrix, $(\text{Tr} \hat{g}_{\kk})^2 \geq 4 \det \hat{g}_{\kk}$. It follows that $\text{Tr} \hat{g}_{\kk} \geq |\Omega_{\kk}|$. Thus we prove Eq.~\eqref{eq:ineqgO}.

When $\text{Tr} \hat{g}_{\kk}$ is equal to its lower bound $|\Omega_{\kk}|$, $\det \hat{g}_{\kk}$ becomes equal to $(\text{Tr} \hat{g}_{\kk})^2/4$, which implies that $g_{\kk}^{-}$ must be 0 if $g_{\kk}^{+}=|\Omega_{\kk}|$.

It is instructive to discuss the quantum geometry in the context of quantum Hall states in the lowest Landau level (LLL). The Berry curvature and quantum metric of magnetic Bloch bands in the LLL is given by
\begin{equation}
|\Omega_{\kk}|= \ell_B^2, \,\,\, \hat{g}_{\kk}=\frac{\ell_B^2}{2}\begin{pmatrix}
1 && 0\\
0 && 1
\end{pmatrix},
\label{LLL}
\end{equation}
which are uniform (i.e., independent of the momentum $\kk$) and saturate the bound $\text{Tr} \hat{g}_{\kk} \geq |\Omega_{\kk}|$. In Eq.~(\ref{LLL}), $\ell_B$ is the magnetic length. With $\Omega_{\kk}$ and $\hat{g}_{\kk}$ in Eq.~(\ref{LLL}), we calculate the spin stiffness using Eq.~\eqref{eq:rhosVOg} [equivalently, Eq.~\eqref{eq:rhosVO}] and find $\rho_s=e^2/(16 \sqrt{2\pi} \epsilon \ell_B)$, which turns out to be the {\it exact} spin stiffness \cite{Yang2006} for the quantum Hall ferromagnetic state in the LLL with Coulomb interaction. Therefore, Eqs.~\eqref{eq:rhosVOg} and \eqref{eq:rhosVO} represent a generalization of spin stiffness from Landau levels to Bloch bands with nonuniform quantum geometry.
 
\bibliographystyle{apsrev4-1}
\bibliography{refs}

\begin{thebibliography}{70}%
\makeatletter
\providecommand \@ifxundefined [1]{%
 \@ifx{#1\undefined}
}%
\providecommand \@ifnum [1]{%
 \ifnum #1\expandafter \@firstoftwo
 \else \expandafter \@secondoftwo
 \fi
}%
\providecommand \@ifx [1]{%
 \ifx #1\expandafter \@firstoftwo
 \else \expandafter \@secondoftwo
 \fi
}%
\providecommand \natexlab [1]{#1}%
\providecommand \enquote  [1]{``#1''}%
\providecommand \bibnamefont  [1]{#1}%
\providecommand \bibfnamefont [1]{#1}%
\providecommand \citenamefont [1]{#1}%
\providecommand \href@noop [0]{\@secondoftwo}%
\providecommand \href [0]{\begingroup \@sanitize@url \@href}%
\providecommand \@href[1]{\@@startlink{#1}\@@href}%
\providecommand \@@href[1]{\endgroup#1\@@endlink}%
\providecommand \@sanitize@url [0]{\catcode `\\12\catcode `\$12\catcode
  `\&12\catcode `\#12\catcode `\^12\catcode `\_12\catcode `\%12\relax}%
\providecommand \@@startlink[1]{}%
\providecommand \@@endlink[0]{}%
\providecommand \url  [0]{\begingroup\@sanitize@url \@url }%
\providecommand \@url [1]{\endgroup\@href {#1}{\urlprefix }}%
\providecommand \urlprefix  [0]{URL }%
\providecommand \Eprint [0]{\href }%
\providecommand \doibase [0]{http://dx.doi.org/}%
\providecommand \selectlanguage [0]{\@gobble}%
\providecommand \bibinfo  [0]{\@secondoftwo}%
\providecommand \bibfield  [0]{\@secondoftwo}%
\providecommand \translation [1]{[#1]}%
\providecommand \BibitemOpen [0]{}%
\providecommand \bibitemStop [0]{}%
\providecommand \bibitemNoStop [0]{.\EOS\space}%
\providecommand \EOS [0]{\spacefactor3000\relax}%
\providecommand \BibitemShut  [1]{\csname bibitem#1\endcsname}%
\let\auto@bib@innerbib\@empty
%</preamble>
\bibitem [{\citenamefont {Bistritzer}\ and\ \citenamefont
  {MacDonald}(2011)}]{Bistritzer2011}%
  \BibitemOpen
  \bibfield  {author} {\bibinfo {author} {\bibfnamefont {R.}~\bibnamefont
  {Bistritzer}}\ and\ \bibinfo {author} {\bibfnamefont {A.~H.}\ \bibnamefont
  {MacDonald}},\ }\href {http://www.pnas.org/content/108/30/12233.abstract}
  {\bibfield  {journal} {\bibinfo  {journal} {Proc. Natl. Acad. Sci. U.S.A.}\
  }\textbf {\bibinfo {volume} {108}},\ \bibinfo {pages} {12233} (\bibinfo
  {year} {2011})}\BibitemShut {NoStop}%
\bibitem [{\citenamefont {Cao}\ \emph {et~al.}(2018{\natexlab{a}})\citenamefont
  {Cao}, \citenamefont {Fatemi}, \citenamefont {Fang}, \citenamefont
  {Watanabe}, \citenamefont {Taniguchi}, \citenamefont {Kaxiras},\ and\
  \citenamefont {Jarillo-Herrero}}]{Cao2018Super}%
  \BibitemOpen
  \bibfield  {author} {\bibinfo {author} {\bibfnamefont {Y.}~\bibnamefont
  {Cao}}, \bibinfo {author} {\bibfnamefont {V.}~\bibnamefont {Fatemi}},
  \bibinfo {author} {\bibfnamefont {S.}~\bibnamefont {Fang}}, \bibinfo {author}
  {\bibfnamefont {K.}~\bibnamefont {Watanabe}}, \bibinfo {author}
  {\bibfnamefont {T.}~\bibnamefont {Taniguchi}}, \bibinfo {author}
  {\bibfnamefont {E.}~\bibnamefont {Kaxiras}}, \ and\ \bibinfo {author}
  {\bibfnamefont {P.}~\bibnamefont {Jarillo-Herrero}},\ }\href
  {http://dx.doi.org/10.1038/nature26160} {\bibfield  {journal} {\bibinfo
  {journal} {Nature}\ }\textbf {\bibinfo {volume} {556}},\ \bibinfo {pages}
  {43} (\bibinfo {year} {2018}{\natexlab{a}})}\BibitemShut {NoStop}%
\bibitem [{\citenamefont {Cao}\ \emph {et~al.}(2018{\natexlab{b}})\citenamefont
  {Cao}, \citenamefont {Fatemi}, \citenamefont {Demir}, \citenamefont {Fang},
  \citenamefont {Tomarken}, \citenamefont {Luo}, \citenamefont
  {Sanchez-Yamagishi}, \citenamefont {Watanabe}, \citenamefont {Taniguchi},
  \citenamefont {Kaxiras}, \citenamefont {Ashoori},\ and\ \citenamefont
  {Jarillo-Herrero}}]{Cao2018Magnetic}%
  \BibitemOpen
  \bibfield  {author} {\bibinfo {author} {\bibfnamefont {Y.}~\bibnamefont
  {Cao}}, \bibinfo {author} {\bibfnamefont {V.}~\bibnamefont {Fatemi}},
  \bibinfo {author} {\bibfnamefont {A.}~\bibnamefont {Demir}}, \bibinfo
  {author} {\bibfnamefont {S.}~\bibnamefont {Fang}}, \bibinfo {author}
  {\bibfnamefont {S.~L.}\ \bibnamefont {Tomarken}}, \bibinfo {author}
  {\bibfnamefont {J.~Y.}\ \bibnamefont {Luo}}, \bibinfo {author} {\bibfnamefont
  {J.~D.}\ \bibnamefont {Sanchez-Yamagishi}}, \bibinfo {author} {\bibfnamefont
  {K.}~\bibnamefont {Watanabe}}, \bibinfo {author} {\bibfnamefont
  {T.}~\bibnamefont {Taniguchi}}, \bibinfo {author} {\bibfnamefont
  {E.}~\bibnamefont {Kaxiras}}, \bibinfo {author} {\bibfnamefont {R.~C.}\
  \bibnamefont {Ashoori}}, \ and\ \bibinfo {author} {\bibfnamefont
  {P.}~\bibnamefont {Jarillo-Herrero}},\ }\href
  {http://dx.doi.org/10.1038/nature26154} {\bibfield  {journal} {\bibinfo
  {journal} {Nature}\ }\textbf {\bibinfo {volume} {556}},\ \bibinfo {pages}
  {80} (\bibinfo {year} {2018}{\natexlab{b}})}\BibitemShut {NoStop}%
\bibitem [{\citenamefont {Yankowitz}\ \emph {et~al.}(2019)\citenamefont
  {Yankowitz}, \citenamefont {Chen}, \citenamefont {Polshyn}, \citenamefont
  {Zhang}, \citenamefont {Watanabe}, \citenamefont {Taniguchi}, \citenamefont
  {Graf}, \citenamefont {Young},\ and\ \citenamefont {Dean}}]{Dean2018tuning}%
  \BibitemOpen
  \bibfield  {author} {\bibinfo {author} {\bibfnamefont {M.}~\bibnamefont
  {Yankowitz}}, \bibinfo {author} {\bibfnamefont {S.}~\bibnamefont {Chen}},
  \bibinfo {author} {\bibfnamefont {H.}~\bibnamefont {Polshyn}}, \bibinfo
  {author} {\bibfnamefont {Y.}~\bibnamefont {Zhang}}, \bibinfo {author}
  {\bibfnamefont {K.}~\bibnamefont {Watanabe}}, \bibinfo {author}
  {\bibfnamefont {T.}~\bibnamefont {Taniguchi}}, \bibinfo {author}
  {\bibfnamefont {D.}~\bibnamefont {Graf}}, \bibinfo {author} {\bibfnamefont
  {A.~F.}\ \bibnamefont {Young}}, \ and\ \bibinfo {author} {\bibfnamefont
  {C.~R.}\ \bibnamefont {Dean}},\ }\href
  {https://science.sciencemag.org/content/363/6431/1059} {\bibfield  {journal}
  {\bibinfo  {journal} {Science}\ }\textbf {\bibinfo {volume} {363}},\ \bibinfo
  {pages} {1059} (\bibinfo {year} {2019})}\BibitemShut {NoStop}%
\bibitem [{\citenamefont {Lu}\ \emph {et~al.}(2019)\citenamefont {Lu},
  \citenamefont {Stepanov}, \citenamefont {Yang}, \citenamefont {Xie},
  \citenamefont {Aamir}, \citenamefont {Das}, \citenamefont {Urgell},
  \citenamefont {Watanabe}, \citenamefont {Taniguchi}, \citenamefont {Zhang}
  \emph {et~al.}}]{lu2019superconductors}%
  \BibitemOpen
  \bibfield  {author} {\bibinfo {author} {\bibfnamefont {X.}~\bibnamefont
  {Lu}}, \bibinfo {author} {\bibfnamefont {P.}~\bibnamefont {Stepanov}},
  \bibinfo {author} {\bibfnamefont {W.}~\bibnamefont {Yang}}, \bibinfo {author}
  {\bibfnamefont {M.}~\bibnamefont {Xie}}, \bibinfo {author} {\bibfnamefont
  {M.~A.}\ \bibnamefont {Aamir}}, \bibinfo {author} {\bibfnamefont
  {I.}~\bibnamefont {Das}}, \bibinfo {author} {\bibfnamefont {C.}~\bibnamefont
  {Urgell}}, \bibinfo {author} {\bibfnamefont {K.}~\bibnamefont {Watanabe}},
  \bibinfo {author} {\bibfnamefont {T.}~\bibnamefont {Taniguchi}}, \bibinfo
  {author} {\bibfnamefont {G.}~\bibnamefont {Zhang}},  \emph {et~al.},\ }\href
  {https://www.nature.com/articles/s41586-019-1695-0} {\bibfield  {journal}
  {\bibinfo  {journal} {Nature}\ }\textbf {\bibinfo {volume} {574}},\ \bibinfo
  {pages} {653} (\bibinfo {year} {2019})}\BibitemShut {NoStop}%
\bibitem [{\citenamefont {Xu}\ and\ \citenamefont
  {Balents}(2018)}]{Balents2018}%
  \BibitemOpen
  \bibfield  {author} {\bibinfo {author} {\bibfnamefont {C.}~\bibnamefont
  {Xu}}\ and\ \bibinfo {author} {\bibfnamefont {L.}~\bibnamefont {Balents}},\
  }\href {\doibase 10.1103/PhysRevLett.121.087001} {\bibfield  {journal}
  {\bibinfo  {journal} {Phys. Rev. Lett.}\ }\textbf {\bibinfo {volume} {121}},\
  \bibinfo {pages} {087001} (\bibinfo {year} {2018})}\BibitemShut {NoStop}%
\bibitem [{\citenamefont {Liu}\ \emph {et~al.}(2018)\citenamefont {Liu},
  \citenamefont {Zhang}, \citenamefont {Chen},\ and\ \citenamefont
  {Yang}}]{Liu2018chiral}%
  \BibitemOpen
  \bibfield  {author} {\bibinfo {author} {\bibfnamefont {C.-C.}\ \bibnamefont
  {Liu}}, \bibinfo {author} {\bibfnamefont {L.-D.}\ \bibnamefont {Zhang}},
  \bibinfo {author} {\bibfnamefont {W.-Q.}\ \bibnamefont {Chen}}, \ and\
  \bibinfo {author} {\bibfnamefont {F.}~\bibnamefont {Yang}},\ }\href {\doibase
  10.1103/PhysRevLett.121.217001} {\bibfield  {journal} {\bibinfo  {journal}
  {Phys. Rev. Lett.}\ }\textbf {\bibinfo {volume} {121}},\ \bibinfo {pages}
  {217001} (\bibinfo {year} {2018})}\BibitemShut {NoStop}%
\bibitem [{\citenamefont {Po}\ \emph {et~al.}(2018)\citenamefont {Po},
  \citenamefont {Zou}, \citenamefont {Vishwanath},\ and\ \citenamefont
  {Senthil}}]{Senthil2018}%
  \BibitemOpen
  \bibfield  {author} {\bibinfo {author} {\bibfnamefont {H.~C.}\ \bibnamefont
  {Po}}, \bibinfo {author} {\bibfnamefont {L.}~\bibnamefont {Zou}}, \bibinfo
  {author} {\bibfnamefont {A.}~\bibnamefont {Vishwanath}}, \ and\ \bibinfo
  {author} {\bibfnamefont {T.}~\bibnamefont {Senthil}},\ }\href {\doibase
  10.1103/PhysRevX.8.031089} {\bibfield  {journal} {\bibinfo  {journal} {Phys.
  Rev. X}\ }\textbf {\bibinfo {volume} {8}},\ \bibinfo {pages} {031089}
  (\bibinfo {year} {2018})}\BibitemShut {NoStop}%
\bibitem [{\citenamefont {Wu}\ \emph {et~al.}(2018)\citenamefont {Wu},
  \citenamefont {MacDonald},\ and\ \citenamefont {Martin}}]{Wu2018phonon}%
  \BibitemOpen
  \bibfield  {author} {\bibinfo {author} {\bibfnamefont {F.}~\bibnamefont
  {Wu}}, \bibinfo {author} {\bibfnamefont {A.~H.}\ \bibnamefont {MacDonald}}, \
  and\ \bibinfo {author} {\bibfnamefont {I.}~\bibnamefont {Martin}},\ }\href
  {\doibase 10.1103/PhysRevLett.121.257001} {\bibfield  {journal} {\bibinfo
  {journal} {Phys. Rev. Lett.}\ }\textbf {\bibinfo {volume} {121}},\ \bibinfo
  {pages} {257001} (\bibinfo {year} {2018})}\BibitemShut {NoStop}%
\bibitem [{\citenamefont {Wu}\ \emph {et~al.}(2019{\natexlab{a}})\citenamefont
  {Wu}, \citenamefont {Hwang},\ and\ \citenamefont {Das~Sarma}}]{wu2019phonon}%
  \BibitemOpen
  \bibfield  {author} {\bibinfo {author} {\bibfnamefont {F.}~\bibnamefont
  {Wu}}, \bibinfo {author} {\bibfnamefont {E.}~\bibnamefont {Hwang}}, \ and\
  \bibinfo {author} {\bibfnamefont {S.}~\bibnamefont {Das~Sarma}},\ }\href
  {\doibase 10.1103/PhysRevB.99.165112} {\bibfield  {journal} {\bibinfo
  {journal} {Phys. Rev. B}\ }\textbf {\bibinfo {volume} {99}},\ \bibinfo
  {pages} {165112} (\bibinfo {year} {2019}{\natexlab{a}})}\BibitemShut
  {NoStop}%
\bibitem [{\citenamefont {Das~Sarma}\ and\ \citenamefont
  {Wu}(2020)}]{sarma2020electron}%
  \BibitemOpen
  \bibfield  {author} {\bibinfo {author} {\bibfnamefont {S.}~\bibnamefont
  {Das~Sarma}}\ and\ \bibinfo {author} {\bibfnamefont {F.}~\bibnamefont {Wu}},\
  }\href {https://doi.org/10.1016/j.aop.2020.168193} {\bibfield  {journal}
  {\bibinfo  {journal} {Annals of Physics}\ ,\ \bibinfo {pages} {168193}}
  (\bibinfo {year} {2020})}\BibitemShut {NoStop}%
\bibitem [{\citenamefont {Peltonen}\ \emph {et~al.}(2018)\citenamefont
  {Peltonen}, \citenamefont {Ojaj\"arvi},\ and\ \citenamefont
  {Heikkil\"a}}]{Heikkila2018}%
  \BibitemOpen
  \bibfield  {author} {\bibinfo {author} {\bibfnamefont {T.~J.}\ \bibnamefont
  {Peltonen}}, \bibinfo {author} {\bibfnamefont {R.}~\bibnamefont
  {Ojaj\"arvi}}, \ and\ \bibinfo {author} {\bibfnamefont {T.~T.}\ \bibnamefont
  {Heikkil\"a}},\ }\href {\doibase 10.1103/PhysRevB.98.220504} {\bibfield
  {journal} {\bibinfo  {journal} {Phys. Rev. B}\ }\textbf {\bibinfo {volume}
  {98}},\ \bibinfo {pages} {220504(R)} (\bibinfo {year} {2018})}\BibitemShut
  {NoStop}%
\bibitem [{\citenamefont {Isobe}\ \emph {et~al.}(2018)\citenamefont {Isobe},
  \citenamefont {Yuan},\ and\ \citenamefont {Fu}}]{Isobe2018}%
  \BibitemOpen
  \bibfield  {author} {\bibinfo {author} {\bibfnamefont {H.}~\bibnamefont
  {Isobe}}, \bibinfo {author} {\bibfnamefont {N.~F.~Q.}\ \bibnamefont {Yuan}},
  \ and\ \bibinfo {author} {\bibfnamefont {L.}~\bibnamefont {Fu}},\ }\href
  {\doibase 10.1103/PhysRevX.8.041041} {\bibfield  {journal} {\bibinfo
  {journal} {Phys. Rev. X}\ }\textbf {\bibinfo {volume} {8}},\ \bibinfo {pages}
  {041041} (\bibinfo {year} {2018})}\BibitemShut {NoStop}%
\bibitem [{\citenamefont {Lian}\ \emph {et~al.}(2019)\citenamefont {Lian},
  \citenamefont {Wang},\ and\ \citenamefont {Bernevig}}]{Lian2018twisted}%
  \BibitemOpen
  \bibfield  {author} {\bibinfo {author} {\bibfnamefont {B.}~\bibnamefont
  {Lian}}, \bibinfo {author} {\bibfnamefont {Z.}~\bibnamefont {Wang}}, \ and\
  \bibinfo {author} {\bibfnamefont {B.~A.}\ \bibnamefont {Bernevig}},\ }\href
  {https://link.aps.org/doi/10.1103/PhysRevLett.122.257002} {\bibfield
  {journal} {\bibinfo  {journal} {Phys. Rev. Lett.}\ }\textbf {\bibinfo
  {volume} {122}},\ \bibinfo {pages} {257002} (\bibinfo {year}
  {2019})}\BibitemShut {NoStop}%
\bibitem [{\citenamefont {Khalaf}\ \emph {et~al.}()\citenamefont {Khalaf},
  \citenamefont {Chatterjee}, \citenamefont {Bultinck}, \citenamefont
  {Zaletel},\ and\ \citenamefont {Vishwanath}}]{khalaf2020charged}%
  \BibitemOpen
  \bibfield  {author} {\bibinfo {author} {\bibfnamefont {E.}~\bibnamefont
  {Khalaf}}, \bibinfo {author} {\bibfnamefont {S.}~\bibnamefont {Chatterjee}},
  \bibinfo {author} {\bibfnamefont {N.}~\bibnamefont {Bultinck}}, \bibinfo
  {author} {\bibfnamefont {M.~P.}\ \bibnamefont {Zaletel}}, \ and\ \bibinfo
  {author} {\bibfnamefont {A.}~\bibnamefont {Vishwanath}},\ }\href
  {https://arxiv.org/abs/2004.00638} {\bibinfo  {journal} {arXiv:2004.00638}\
  }\BibitemShut {NoStop}%
\bibitem [{\citenamefont {Koshino}\ \emph {et~al.}(2018)\citenamefont
  {Koshino}, \citenamefont {Yuan}, \citenamefont {Koretsune}, \citenamefont
  {Ochi}, \citenamefont {Kuroki},\ and\ \citenamefont {Fu}}]{Koshino2018}%
  \BibitemOpen
\bibfield  {journal} {  }\bibfield  {author} {\bibinfo {author} {\bibfnamefont
  {M.}~\bibnamefont {Koshino}}, \bibinfo {author} {\bibfnamefont {N.~F.~Q.}\
  \bibnamefont {Yuan}}, \bibinfo {author} {\bibfnamefont {T.}~\bibnamefont
  {Koretsune}}, \bibinfo {author} {\bibfnamefont {M.}~\bibnamefont {Ochi}},
  \bibinfo {author} {\bibfnamefont {K.}~\bibnamefont {Kuroki}}, \ and\ \bibinfo
  {author} {\bibfnamefont {L.}~\bibnamefont {Fu}},\ }\href {\doibase
  10.1103/PhysRevX.8.031087} {\bibfield  {journal} {\bibinfo  {journal} {Phys.
  Rev. X}\ }\textbf {\bibinfo {volume} {8}},\ \bibinfo {pages} {031087}
  (\bibinfo {year} {2018})}\BibitemShut {NoStop}%
\bibitem [{\citenamefont {Kang}\ and\ \citenamefont {Vafek}(2018)}]{Kang2018}%
  \BibitemOpen
  \bibfield  {author} {\bibinfo {author} {\bibfnamefont {J.}~\bibnamefont
  {Kang}}\ and\ \bibinfo {author} {\bibfnamefont {O.}~\bibnamefont {Vafek}},\
  }\href {\doibase 10.1103/PhysRevX.8.031088} {\bibfield  {journal} {\bibinfo
  {journal} {Phys. Rev. X}\ }\textbf {\bibinfo {volume} {8}},\ \bibinfo {pages}
  {031088} (\bibinfo {year} {2018})}\BibitemShut {NoStop}%
\bibitem [{\citenamefont {Rademaker}\ \emph {et~al.}(2019)\citenamefont
  {Rademaker}, \citenamefont {Abanin},\ and\ \citenamefont
  {Mellado}}]{Rademaker2019}%
  \BibitemOpen
  \bibfield  {author} {\bibinfo {author} {\bibfnamefont {L.}~\bibnamefont
  {Rademaker}}, \bibinfo {author} {\bibfnamefont {D.~A.}\ \bibnamefont
  {Abanin}}, \ and\ \bibinfo {author} {\bibfnamefont {P.}~\bibnamefont
  {Mellado}},\ }\href {\doibase 10.1103/PhysRevB.100.205114} {\bibfield
  {journal} {\bibinfo  {journal} {Phys. Rev. B}\ }\textbf {\bibinfo {volume}
  {100}},\ \bibinfo {pages} {205114} (\bibinfo {year} {2019})}\BibitemShut
  {NoStop}%
\bibitem [{\citenamefont {Bultinck}\ \emph {et~al.}()\citenamefont {Bultinck},
  \citenamefont {Khalaf}, \citenamefont {Liu}, \citenamefont {Chatterjee},
  \citenamefont {Vishwanath},\ and\ \citenamefont
  {Zaletel}}]{bultinck2019ground}%
  \BibitemOpen
  \bibfield  {author} {\bibinfo {author} {\bibfnamefont {N.}~\bibnamefont
  {Bultinck}}, \bibinfo {author} {\bibfnamefont {E.}~\bibnamefont {Khalaf}},
  \bibinfo {author} {\bibfnamefont {S.}~\bibnamefont {Liu}}, \bibinfo {author}
  {\bibfnamefont {S.}~\bibnamefont {Chatterjee}}, \bibinfo {author}
  {\bibfnamefont {A.}~\bibnamefont {Vishwanath}}, \ and\ \bibinfo {author}
  {\bibfnamefont {M.~P.}\ \bibnamefont {Zaletel}},\ }\href
  {https://arxiv.org/abs/1911.02045} {\bibinfo  {journal} {arXiv:1911.02045}\
  }\BibitemShut {NoStop}%
\bibitem [{\citenamefont {Zhang}\ \emph {et~al.}()\citenamefont {Zhang},
  \citenamefont {Jiang}, \citenamefont {Wang},\ and\ \citenamefont
  {Zhang}}]{zhang2020correlated}%
  \BibitemOpen
\bibfield  {journal} {  }\bibfield  {author} {\bibinfo {author} {\bibfnamefont
  {Y.}~\bibnamefont {Zhang}}, \bibinfo {author} {\bibfnamefont
  {K.}~\bibnamefont {Jiang}}, \bibinfo {author} {\bibfnamefont
  {Z.}~\bibnamefont {Wang}}, \ and\ \bibinfo {author} {\bibfnamefont
  {F.}~\bibnamefont {Zhang}},\ }\href {https://arxiv.org/abs/2001.02476}
  {\bibinfo  {journal} {arXiv:2001.02476}\ }\BibitemShut {NoStop}%
\bibitem [{\citenamefont {Kang}\ and\ \citenamefont {Vafek}()}]{kang2020non}%
  \BibitemOpen
\bibfield  {journal} {  }\bibfield  {author} {\bibinfo {author} {\bibfnamefont
  {J.}~\bibnamefont {Kang}}\ and\ \bibinfo {author} {\bibfnamefont
  {O.}~\bibnamefont {Vafek}},\ }\href {https://arxiv.org/abs/2002.10360}
  {\bibinfo  {journal} {arXiv:2002.10360}\ }\BibitemShut {NoStop}%
\bibitem [{\citenamefont {Hsu}\ \emph {et~al.}()\citenamefont {Hsu},
  \citenamefont {Wu},\ and\ \citenamefont {Sarma}}]{hsu2020topological}%
  \BibitemOpen
\bibfield  {journal} {  }\bibfield  {author} {\bibinfo {author} {\bibfnamefont
  {Y.-T.}\ \bibnamefont {Hsu}}, \bibinfo {author} {\bibfnamefont
  {F.}~\bibnamefont {Wu}}, \ and\ \bibinfo {author} {\bibfnamefont {S.~D.}\
  \bibnamefont {Sarma}},\ }\href {https://arxiv.org/abs/2003.02847} {\bibinfo
  {journal} {arXiv:2003.02847}\ }\BibitemShut {NoStop}%
\bibitem [{\citenamefont {Cea}\ and\ \citenamefont {Guinea}()}]{cea2020band}%
  \BibitemOpen
\bibfield  {journal} {  }\bibfield  {author} {\bibinfo {author} {\bibfnamefont
  {T.}~\bibnamefont {Cea}}\ and\ \bibinfo {author} {\bibfnamefont
  {F.}~\bibnamefont {Guinea}},\ }\href {https://arxiv.org/abs/2004.01577}
  {\bibinfo  {journal} {arXiv:2004.01577}\ }\BibitemShut {NoStop}%
\bibitem [{\citenamefont {Zhang}\ \emph
  {et~al.}(2019{\natexlab{a}})\citenamefont {Zhang}, \citenamefont {Mao},
  \citenamefont {Cao}, \citenamefont {Jarillo-Herrero},\ and\ \citenamefont
  {Senthil}}]{Zhang2019}%
  \BibitemOpen
\bibfield  {journal} {  }\bibfield  {author} {\bibinfo {author} {\bibfnamefont
  {Y.-H.}\ \bibnamefont {Zhang}}, \bibinfo {author} {\bibfnamefont
  {D.}~\bibnamefont {Mao}}, \bibinfo {author} {\bibfnamefont {Y.}~\bibnamefont
  {Cao}}, \bibinfo {author} {\bibfnamefont {P.}~\bibnamefont
  {Jarillo-Herrero}}, \ and\ \bibinfo {author} {\bibfnamefont {T.}~\bibnamefont
  {Senthil}},\ }\href {\doibase 10.1103/PhysRevB.99.075127} {\bibfield
  {journal} {\bibinfo  {journal} {Phys. Rev. B}\ }\textbf {\bibinfo {volume}
  {99}},\ \bibinfo {pages} {075127} (\bibinfo {year}
  {2019}{\natexlab{a}})}\BibitemShut {NoStop}%
\bibitem [{\citenamefont {Xie}\ and\ \citenamefont
  {MacDonald}(2020)}]{xie2018nature}%
  \BibitemOpen
  \bibfield  {author} {\bibinfo {author} {\bibfnamefont {M.}~\bibnamefont
  {Xie}}\ and\ \bibinfo {author} {\bibfnamefont {A.~H.}\ \bibnamefont
  {MacDonald}},\ }\href {\doibase 10.1103/PhysRevLett.124.097601} {\bibfield
  {journal} {\bibinfo  {journal} {Phys. Rev. Lett.}\ }\textbf {\bibinfo
  {volume} {124}},\ \bibinfo {pages} {097601} (\bibinfo {year}
  {2020})}\BibitemShut {NoStop}%
\bibitem [{\citenamefont {Kang}\ and\ \citenamefont
  {Vafek}(2019)}]{kang2018strong}%
  \BibitemOpen
  \bibfield  {author} {\bibinfo {author} {\bibfnamefont {J.}~\bibnamefont
  {Kang}}\ and\ \bibinfo {author} {\bibfnamefont {O.}~\bibnamefont {Vafek}},\
  }\href {\doibase 10.1103/PhysRevLett.122.246401} {\bibfield  {journal}
  {\bibinfo  {journal} {Phys. Rev. Lett.}\ }\textbf {\bibinfo {volume} {122}},\
  \bibinfo {pages} {246401} (\bibinfo {year} {2019})}\BibitemShut {NoStop}%
\bibitem [{\citenamefont {Seo}\ \emph {et~al.}(2019)\citenamefont {Seo},
  \citenamefont {Kotov},\ and\ \citenamefont {Uchoa}}]{Seo2019}%
  \BibitemOpen
  \bibfield  {author} {\bibinfo {author} {\bibfnamefont {K.}~\bibnamefont
  {Seo}}, \bibinfo {author} {\bibfnamefont {V.~N.}\ \bibnamefont {Kotov}}, \
  and\ \bibinfo {author} {\bibfnamefont {B.}~\bibnamefont {Uchoa}},\ }\href
  {\doibase 10.1103/PhysRevLett.122.246402} {\bibfield  {journal} {\bibinfo
  {journal} {Phys. Rev. Lett.}\ }\textbf {\bibinfo {volume} {122}},\ \bibinfo
  {pages} {246402} (\bibinfo {year} {2019})}\BibitemShut {NoStop}%
\bibitem [{\citenamefont {Wu}\ \emph {et~al.}(2019{\natexlab{b}})\citenamefont
  {Wu}, \citenamefont {Lovorn}, \citenamefont {Tutuc}, \citenamefont {Martin},\
  and\ \citenamefont {MacDonald}}]{WuTITMD}%
  \BibitemOpen
  \bibfield  {author} {\bibinfo {author} {\bibfnamefont {F.}~\bibnamefont
  {Wu}}, \bibinfo {author} {\bibfnamefont {T.}~\bibnamefont {Lovorn}}, \bibinfo
  {author} {\bibfnamefont {E.}~\bibnamefont {Tutuc}}, \bibinfo {author}
  {\bibfnamefont {I.}~\bibnamefont {Martin}}, \ and\ \bibinfo {author}
  {\bibfnamefont {A.~H.}\ \bibnamefont {MacDonald}},\ }\href {\doibase
  10.1103/PhysRevLett.122.086402} {\bibfield  {journal} {\bibinfo  {journal}
  {Phys. Rev. Lett.}\ }\textbf {\bibinfo {volume} {122}},\ \bibinfo {pages}
  {086402} (\bibinfo {year} {2019}{\natexlab{b}})}\BibitemShut {NoStop}%
\bibitem [{\citenamefont {Liu}\ \emph {et~al.}(2019)\citenamefont {Liu},
  \citenamefont {Ma}, \citenamefont {Gao},\ and\ \citenamefont
  {Dai}}]{LiuMulti}%
  \BibitemOpen
  \bibfield  {author} {\bibinfo {author} {\bibfnamefont {J.}~\bibnamefont
  {Liu}}, \bibinfo {author} {\bibfnamefont {Z.}~\bibnamefont {Ma}}, \bibinfo
  {author} {\bibfnamefont {J.}~\bibnamefont {Gao}}, \ and\ \bibinfo {author}
  {\bibfnamefont {X.}~\bibnamefont {Dai}},\ }\href {\doibase
  10.1103/PhysRevX.9.031021} {\bibfield  {journal} {\bibinfo  {journal} {Phys.
  Rev. X}\ }\textbf {\bibinfo {volume} {9}},\ \bibinfo {pages} {031021}
  (\bibinfo {year} {2019})}\BibitemShut {NoStop}%
\bibitem [{\citenamefont {Wolf}\ \emph {et~al.}(2019)\citenamefont {Wolf},
  \citenamefont {Lado}, \citenamefont {Blatter},\ and\ \citenamefont
  {Zilberberg}}]{Wolf2019}%
  \BibitemOpen
  \bibfield  {author} {\bibinfo {author} {\bibfnamefont {T.~M.~R.}\
  \bibnamefont {Wolf}}, \bibinfo {author} {\bibfnamefont {J.~L.}\ \bibnamefont
  {Lado}}, \bibinfo {author} {\bibfnamefont {G.}~\bibnamefont {Blatter}}, \
  and\ \bibinfo {author} {\bibfnamefont {O.}~\bibnamefont {Zilberberg}},\
  }\href {\doibase 10.1103/PhysRevLett.123.096802} {\bibfield  {journal}
  {\bibinfo  {journal} {Phys. Rev. Lett.}\ }\textbf {\bibinfo {volume} {123}},\
  \bibinfo {pages} {096802} (\bibinfo {year} {2019})}\BibitemShut {NoStop}%
\bibitem [{\citenamefont {Sharpe}\ \emph {et~al.}(2019)\citenamefont {Sharpe},
  \citenamefont {Fox}, \citenamefont {Barnard}, \citenamefont {Finney},
  \citenamefont {Watanabe}, \citenamefont {Taniguchi}, \citenamefont
  {Kastner},\ and\ \citenamefont {Goldhaber-Gordon}}]{sharpe2019emergent}%
  \BibitemOpen
  \bibfield  {author} {\bibinfo {author} {\bibfnamefont {A.~L.}\ \bibnamefont
  {Sharpe}}, \bibinfo {author} {\bibfnamefont {E.~J.}\ \bibnamefont {Fox}},
  \bibinfo {author} {\bibfnamefont {A.~W.}\ \bibnamefont {Barnard}}, \bibinfo
  {author} {\bibfnamefont {J.}~\bibnamefont {Finney}}, \bibinfo {author}
  {\bibfnamefont {K.}~\bibnamefont {Watanabe}}, \bibinfo {author}
  {\bibfnamefont {T.}~\bibnamefont {Taniguchi}}, \bibinfo {author}
  {\bibfnamefont {M.~A.}\ \bibnamefont {Kastner}}, \ and\ \bibinfo {author}
  {\bibfnamefont {D.}~\bibnamefont {Goldhaber-Gordon}},\ }\href {\doibase
  10.1126/science.aaw3780} {\bibfield  {journal} {\bibinfo  {journal}
  {Science}\ }\textbf {\bibinfo {volume} {365}},\ \bibinfo {pages} {605}
  (\bibinfo {year} {2019})}\BibitemShut {NoStop}%
\bibitem [{\citenamefont {Serlin}\ \emph {et~al.}(2020)\citenamefont {Serlin},
  \citenamefont {Tschirhart}, \citenamefont {Polshyn}, \citenamefont {Zhang},
  \citenamefont {Zhu}, \citenamefont {Watanabe}, \citenamefont {Taniguchi},
  \citenamefont {Balents},\ and\ \citenamefont {Young}}]{serlin2020intrinsic}%
  \BibitemOpen
  \bibfield  {author} {\bibinfo {author} {\bibfnamefont {M.}~\bibnamefont
  {Serlin}}, \bibinfo {author} {\bibfnamefont {C.}~\bibnamefont {Tschirhart}},
  \bibinfo {author} {\bibfnamefont {H.}~\bibnamefont {Polshyn}}, \bibinfo
  {author} {\bibfnamefont {Y.}~\bibnamefont {Zhang}}, \bibinfo {author}
  {\bibfnamefont {J.}~\bibnamefont {Zhu}}, \bibinfo {author} {\bibfnamefont
  {K.}~\bibnamefont {Watanabe}}, \bibinfo {author} {\bibfnamefont
  {T.}~\bibnamefont {Taniguchi}}, \bibinfo {author} {\bibfnamefont
  {L.}~\bibnamefont {Balents}}, \ and\ \bibinfo {author} {\bibfnamefont
  {A.}~\bibnamefont {Young}},\ }\href
  {https://science.sciencemag.org/content/367/6480/900} {\bibfield  {journal}
  {\bibinfo  {journal} {Science}\ }\textbf {\bibinfo {volume} {367}},\ \bibinfo
  {pages} {900} (\bibinfo {year} {2020})}\BibitemShut {NoStop}%
\bibitem [{\citenamefont {Shen}\ \emph {et~al.}(2020)\citenamefont {Shen},
  \citenamefont {Chu}, \citenamefont {Wu}, \citenamefont {Li}, \citenamefont
  {Wang}, \citenamefont {Zhao}, \citenamefont {Tang}, \citenamefont {Liu},
  \citenamefont {Tian}, \citenamefont {Watanabe}, \citenamefont {Taniguchi},
  \citenamefont {Yang}, \citenamefont {Meng}, \citenamefont {Shi},
  \citenamefont {Yazyev},\ and\ \citenamefont {Zhang}}]{Shen2020}%
  \BibitemOpen
  \bibfield  {author} {\bibinfo {author} {\bibfnamefont {C.}~\bibnamefont
  {Shen}}, \bibinfo {author} {\bibfnamefont {Y.}~\bibnamefont {Chu}}, \bibinfo
  {author} {\bibfnamefont {Q.}~\bibnamefont {Wu}}, \bibinfo {author}
  {\bibfnamefont {N.}~\bibnamefont {Li}}, \bibinfo {author} {\bibfnamefont
  {S.}~\bibnamefont {Wang}}, \bibinfo {author} {\bibfnamefont {Y.}~\bibnamefont
  {Zhao}}, \bibinfo {author} {\bibfnamefont {J.}~\bibnamefont {Tang}}, \bibinfo
  {author} {\bibfnamefont {J.}~\bibnamefont {Liu}}, \bibinfo {author}
  {\bibfnamefont {J.}~\bibnamefont {Tian}}, \bibinfo {author} {\bibfnamefont
  {K.}~\bibnamefont {Watanabe}}, \bibinfo {author} {\bibfnamefont
  {T.}~\bibnamefont {Taniguchi}}, \bibinfo {author} {\bibfnamefont
  {R.}~\bibnamefont {Yang}}, \bibinfo {author} {\bibfnamefont {Z.~Y.}\
  \bibnamefont {Meng}}, \bibinfo {author} {\bibfnamefont {D.}~\bibnamefont
  {Shi}}, \bibinfo {author} {\bibfnamefont {O.~V.}\ \bibnamefont {Yazyev}}, \
  and\ \bibinfo {author} {\bibfnamefont {G.}~\bibnamefont {Zhang}},\ }\href
  {https://doi.org/10.1038/s41567-020-0825-9} {\bibfield  {journal} {\bibinfo
  {journal} {Nature Physics}\ }\textbf {\bibinfo {volume} {16}},\ \bibinfo
  {pages} {520} (\bibinfo {year} {2020})}\BibitemShut {NoStop}%
\bibitem [{\citenamefont {Liu}\ \emph {et~al.}()\citenamefont {Liu},
  \citenamefont {Hao}, \citenamefont {Khalaf}, \citenamefont {Lee},
  \citenamefont {Watanabe}, \citenamefont {Taniguchi}, \citenamefont
  {Vishwanath},\ and\ \citenamefont {Kim}}]{liu2019spin}%
  \BibitemOpen
  \bibfield  {author} {\bibinfo {author} {\bibfnamefont {X.}~\bibnamefont
  {Liu}}, \bibinfo {author} {\bibfnamefont {Z.}~\bibnamefont {Hao}}, \bibinfo
  {author} {\bibfnamefont {E.}~\bibnamefont {Khalaf}}, \bibinfo {author}
  {\bibfnamefont {J.~Y.}\ \bibnamefont {Lee}}, \bibinfo {author} {\bibfnamefont
  {K.}~\bibnamefont {Watanabe}}, \bibinfo {author} {\bibfnamefont
  {T.}~\bibnamefont {Taniguchi}}, \bibinfo {author} {\bibfnamefont
  {A.}~\bibnamefont {Vishwanath}}, \ and\ \bibinfo {author} {\bibfnamefont
  {P.}~\bibnamefont {Kim}},\ }\href {https://arxiv.org/abs/1903.08130}
  {\bibinfo  {journal} {arXiv:1903.08130}\ }\BibitemShut {NoStop}%
\bibitem [{\citenamefont {Cao}\ \emph {et~al.}(2020)\citenamefont {Cao},
  \citenamefont {Rodan-Legrain}, \citenamefont {Rubies-Bigorda}, \citenamefont
  {Park}, \citenamefont {Watanabe}, \citenamefont {Taniguchi},\ and\
  \citenamefont {Jarillo-Herrero}}]{cao2020tunable}%
  \BibitemOpen
\bibfield  {journal} {  }\bibfield  {author} {\bibinfo {author} {\bibfnamefont
  {Y.}~\bibnamefont {Cao}}, \bibinfo {author} {\bibfnamefont {D.}~\bibnamefont
  {Rodan-Legrain}}, \bibinfo {author} {\bibfnamefont {O.}~\bibnamefont
  {Rubies-Bigorda}}, \bibinfo {author} {\bibfnamefont {J.~M.}\ \bibnamefont
  {Park}}, \bibinfo {author} {\bibfnamefont {K.}~\bibnamefont {Watanabe}},
  \bibinfo {author} {\bibfnamefont {T.}~\bibnamefont {Taniguchi}}, \ and\
  \bibinfo {author} {\bibfnamefont {P.}~\bibnamefont {Jarillo-Herrero}},\
  }\href {https://www.nature.com/articles/s41586-020-2260-6?draft=collection}
  {\bibfield  {journal} {\bibinfo  {journal} {Nature}\ } (\bibinfo {year}
  {2020})}\BibitemShut {NoStop}%
\bibitem [{\citenamefont {Burg}\ \emph {et~al.}(2019)\citenamefont {Burg},
  \citenamefont {Zhu}, \citenamefont {Taniguchi}, \citenamefont {Watanabe},
  \citenamefont {MacDonald},\ and\ \citenamefont {Tutuc}}]{Burg2019}%
  \BibitemOpen
  \bibfield  {author} {\bibinfo {author} {\bibfnamefont {G.~W.}\ \bibnamefont
  {Burg}}, \bibinfo {author} {\bibfnamefont {J.}~\bibnamefont {Zhu}}, \bibinfo
  {author} {\bibfnamefont {T.}~\bibnamefont {Taniguchi}}, \bibinfo {author}
  {\bibfnamefont {K.}~\bibnamefont {Watanabe}}, \bibinfo {author}
  {\bibfnamefont {A.~H.}\ \bibnamefont {MacDonald}}, \ and\ \bibinfo {author}
  {\bibfnamefont {E.}~\bibnamefont {Tutuc}},\ }\href {\doibase
  10.1103/PhysRevLett.123.197702} {\bibfield  {journal} {\bibinfo  {journal}
  {Phys. Rev. Lett.}\ }\textbf {\bibinfo {volume} {123}},\ \bibinfo {pages}
  {197702} (\bibinfo {year} {2019})}\BibitemShut {NoStop}%
\bibitem [{\citenamefont {Chen}\ \emph {et~al.}(2020)\citenamefont {Chen},
  \citenamefont {Sharpe}, \citenamefont {Fox}, \citenamefont {Zhang},
  \citenamefont {Wang}, \citenamefont {Jiang}, \citenamefont {Lyu},
  \citenamefont {Li}, \citenamefont {Watanabe}, \citenamefont {Taniguchi} \emph
  {et~al.}}]{chen2020tunable}%
  \BibitemOpen
  \bibfield  {author} {\bibinfo {author} {\bibfnamefont {G.}~\bibnamefont
  {Chen}}, \bibinfo {author} {\bibfnamefont {A.~L.}\ \bibnamefont {Sharpe}},
  \bibinfo {author} {\bibfnamefont {E.~J.}\ \bibnamefont {Fox}}, \bibinfo
  {author} {\bibfnamefont {Y.-H.}\ \bibnamefont {Zhang}}, \bibinfo {author}
  {\bibfnamefont {S.}~\bibnamefont {Wang}}, \bibinfo {author} {\bibfnamefont
  {L.}~\bibnamefont {Jiang}}, \bibinfo {author} {\bibfnamefont
  {B.}~\bibnamefont {Lyu}}, \bibinfo {author} {\bibfnamefont {H.}~\bibnamefont
  {Li}}, \bibinfo {author} {\bibfnamefont {K.}~\bibnamefont {Watanabe}},
  \bibinfo {author} {\bibfnamefont {T.}~\bibnamefont {Taniguchi}},  \emph
  {et~al.},\ }\href {https://www.nature.com/articles/s41586-020-2049-7}
  {\bibfield  {journal} {\bibinfo  {journal} {Nature}\ }\textbf {\bibinfo
  {volume} {579}},\ \bibinfo {pages} {56} (\bibinfo {year} {2020})}\BibitemShut
  {NoStop}%
\bibitem [{\citenamefont {Polshyn}\ \emph {et~al.}()\citenamefont {Polshyn},
  \citenamefont {Zhu}, \citenamefont {Kumar}, \citenamefont {Zhang},
  \citenamefont {Yang}, \citenamefont {Tschirhart}, \citenamefont {Serlin},
  \citenamefont {Watanabe}, \citenamefont {Taniguchi}, \citenamefont
  {MacDonald} \emph {et~al.}}]{polshyn2020nonvolatile}%
  \BibitemOpen
  \bibfield  {author} {\bibinfo {author} {\bibfnamefont {H.}~\bibnamefont
  {Polshyn}}, \bibinfo {author} {\bibfnamefont {J.}~\bibnamefont {Zhu}},
  \bibinfo {author} {\bibfnamefont {M.~A.}\ \bibnamefont {Kumar}}, \bibinfo
  {author} {\bibfnamefont {Y.}~\bibnamefont {Zhang}}, \bibinfo {author}
  {\bibfnamefont {F.}~\bibnamefont {Yang}}, \bibinfo {author} {\bibfnamefont
  {C.~L.}\ \bibnamefont {Tschirhart}}, \bibinfo {author} {\bibfnamefont
  {M.}~\bibnamefont {Serlin}}, \bibinfo {author} {\bibfnamefont
  {K.}~\bibnamefont {Watanabe}}, \bibinfo {author} {\bibfnamefont
  {T.}~\bibnamefont {Taniguchi}}, \bibinfo {author} {\bibfnamefont {A.~H.}\
  \bibnamefont {MacDonald}},  \emph {et~al.},\ }\href
  {https://arxiv.org/abs/2004.11353} {\bibinfo  {journal} {arXiv:2004.11353}\
  }\BibitemShut {NoStop}%
\bibitem [{\citenamefont {Chen}\ \emph {et~al.}()\citenamefont {Chen},
  \citenamefont {He}, \citenamefont {Zhang}, \citenamefont {Hsieh},
  \citenamefont {Fei}, \citenamefont {Watanabe}, \citenamefont {Taniguchi},
  \citenamefont {Cobden}, \citenamefont {Xu}, \citenamefont {Dean} \emph
  {et~al.}}]{chen2020electrically}%
  \BibitemOpen
\bibfield  {journal} {  }\bibfield  {author} {\bibinfo {author} {\bibfnamefont
  {S.}~\bibnamefont {Chen}}, \bibinfo {author} {\bibfnamefont {M.}~\bibnamefont
  {He}}, \bibinfo {author} {\bibfnamefont {Y.-H.}\ \bibnamefont {Zhang}},
  \bibinfo {author} {\bibfnamefont {V.}~\bibnamefont {Hsieh}}, \bibinfo
  {author} {\bibfnamefont {Z.}~\bibnamefont {Fei}}, \bibinfo {author}
  {\bibfnamefont {K.}~\bibnamefont {Watanabe}}, \bibinfo {author}
  {\bibfnamefont {T.}~\bibnamefont {Taniguchi}}, \bibinfo {author}
  {\bibfnamefont {D.~H.}\ \bibnamefont {Cobden}}, \bibinfo {author}
  {\bibfnamefont {X.}~\bibnamefont {Xu}}, \bibinfo {author} {\bibfnamefont
  {C.~R.}\ \bibnamefont {Dean}},  \emph {et~al.},\ }\href
  {https://arxiv.org/abs/2004.11340} {\bibinfo  {journal} {arXiv:2004.11340}\
  }\BibitemShut {NoStop}%
\bibitem [{\citenamefont {Zhang}\ \emph
  {et~al.}(2019{\natexlab{b}})\citenamefont {Zhang}, \citenamefont {Mao},\ and\
  \citenamefont {Senthil}}]{Zhang2019Twisted}%
  \BibitemOpen
\bibfield  {journal} {  }\bibfield  {author} {\bibinfo {author} {\bibfnamefont
  {Y.-H.}\ \bibnamefont {Zhang}}, \bibinfo {author} {\bibfnamefont
  {D.}~\bibnamefont {Mao}}, \ and\ \bibinfo {author} {\bibfnamefont
  {T.}~\bibnamefont {Senthil}},\ }\href {\doibase
  10.1103/PhysRevResearch.1.033126} {\bibfield  {journal} {\bibinfo  {journal}
  {Phys. Rev. Research}\ }\textbf {\bibinfo {volume} {1}},\ \bibinfo {pages}
  {033126} (\bibinfo {year} {2019}{\natexlab{b}})}\BibitemShut {NoStop}%
\bibitem [{\citenamefont {Bultinck}\ \emph {et~al.}(2020)\citenamefont
  {Bultinck}, \citenamefont {Chatterjee},\ and\ \citenamefont
  {Zaletel}}]{bultinck2019anomalous}%
  \BibitemOpen
  \bibfield  {author} {\bibinfo {author} {\bibfnamefont {N.}~\bibnamefont
  {Bultinck}}, \bibinfo {author} {\bibfnamefont {S.}~\bibnamefont
  {Chatterjee}}, \ and\ \bibinfo {author} {\bibfnamefont {M.~P.}\ \bibnamefont
  {Zaletel}},\ }\href {\doibase 10.1103/PhysRevLett.124.166601} {\bibfield
  {journal} {\bibinfo  {journal} {Phys. Rev. Lett.}\ }\textbf {\bibinfo
  {volume} {124}},\ \bibinfo {pages} {166601} (\bibinfo {year}
  {2020})}\BibitemShut {NoStop}%
\bibitem [{\citenamefont {Repellin}\ \emph {et~al.}(2020)\citenamefont
  {Repellin}, \citenamefont {Dong}, \citenamefont {Zhang},\ and\ \citenamefont
  {Senthil}}]{repellin2019ferromagnetism}%
  \BibitemOpen
  \bibfield  {author} {\bibinfo {author} {\bibfnamefont {C.}~\bibnamefont
  {Repellin}}, \bibinfo {author} {\bibfnamefont {Z.}~\bibnamefont {Dong}},
  \bibinfo {author} {\bibfnamefont {Y.-H.}\ \bibnamefont {Zhang}}, \ and\
  \bibinfo {author} {\bibfnamefont {T.}~\bibnamefont {Senthil}},\ }\href
  {\doibase 10.1103/PhysRevLett.124.187601} {\bibfield  {journal} {\bibinfo
  {journal} {Phys. Rev. Lett.}\ }\textbf {\bibinfo {volume} {124}},\ \bibinfo
  {pages} {187601} (\bibinfo {year} {2020})}\BibitemShut {NoStop}%
\bibitem [{\citenamefont {Alavirad}\ and\ \citenamefont
  {Sau}()}]{alavirad2019ferromagnetism}%
  \BibitemOpen
  \bibfield  {author} {\bibinfo {author} {\bibfnamefont {Y.}~\bibnamefont
  {Alavirad}}\ and\ \bibinfo {author} {\bibfnamefont {J.~D.}\ \bibnamefont
  {Sau}},\ }\href {https://arxiv.org/abs/1907.13633} {\bibinfo  {journal}
  {arXiv:1907.13633}\ }\BibitemShut {NoStop}%
\bibitem [{\citenamefont {Wu}\ and\ \citenamefont
  {Das~Sarma}(2020{\natexlab{a}})}]{Wu2020Collective}%
  \BibitemOpen
\bibfield  {journal} {  }\bibfield  {author} {\bibinfo {author} {\bibfnamefont
  {F.}~\bibnamefont {Wu}}\ and\ \bibinfo {author} {\bibfnamefont
  {S.}~\bibnamefont {Das~Sarma}},\ }\href {\doibase
  10.1103/PhysRevLett.124.046403} {\bibfield  {journal} {\bibinfo  {journal}
  {Phys. Rev. Lett.}\ }\textbf {\bibinfo {volume} {124}},\ \bibinfo {pages}
  {046403} (\bibinfo {year} {2020}{\natexlab{a}})}\BibitemShut {NoStop}%
\bibitem [{\citenamefont {Liu}\ and\ \citenamefont
  {Dai}({\natexlab{a}})}]{liu2019correlated}%
  \BibitemOpen
  \bibfield  {author} {\bibinfo {author} {\bibfnamefont {J.}~\bibnamefont
  {Liu}}\ and\ \bibinfo {author} {\bibfnamefont {X.}~\bibnamefont {Dai}},\
  }\href {https://arxiv.org/abs/1911.03760} {\bibfield  {journal} {\bibinfo
  {journal} {arXiv:1911.03760}\ } ({\natexlab{a}})}\BibitemShut {NoStop}%
\bibitem [{\citenamefont {Liu}\ and\ \citenamefont
  {Dai}({\natexlab{b}})}]{liu2019anomalous}%
  \BibitemOpen
  \bibfield  {author} {\bibinfo {author} {\bibfnamefont {J.}~\bibnamefont
  {Liu}}\ and\ \bibinfo {author} {\bibfnamefont {X.}~\bibnamefont {Dai}},\
  }\href {https://arxiv.org/abs/1907.08932} {\bibfield  {journal} {\bibinfo
  {journal} {arXiv:1907.08932}\ } ({\natexlab{b}})}\BibitemShut {NoStop}%
\bibitem [{\citenamefont {He}\ \emph {et~al.}(2020)\citenamefont {He},
  \citenamefont {Goldhaber-Gordon},\ and\ \citenamefont {Law}}]{He2020}%
  \BibitemOpen
  \bibfield  {author} {\bibinfo {author} {\bibfnamefont {W.-Y.}\ \bibnamefont
  {He}}, \bibinfo {author} {\bibfnamefont {D.}~\bibnamefont
  {Goldhaber-Gordon}}, \ and\ \bibinfo {author} {\bibfnamefont {K.~T.}\
  \bibnamefont {Law}},\ }\href {https://doi.org/10.1038/s41467-020-15473-9}
  {\bibfield  {journal} {\bibinfo  {journal} {Nat. Commun.}\ }\textbf {\bibinfo
  {volume} {11}},\ \bibinfo {pages} {1650} (\bibinfo {year}
  {2020})}\BibitemShut {NoStop}%
\bibitem [{\citenamefont {Zhu}\ \emph {et~al.}()\citenamefont {Zhu},
  \citenamefont {Su},\ and\ \citenamefont {MacDonald}}]{zhu2020curious}%
  \BibitemOpen
  \bibfield  {author} {\bibinfo {author} {\bibfnamefont {J.}~\bibnamefont
  {Zhu}}, \bibinfo {author} {\bibfnamefont {J.-J.}\ \bibnamefont {Su}}, \ and\
  \bibinfo {author} {\bibfnamefont {A.~H.}\ \bibnamefont {MacDonald}},\ }\href
  {https://arxiv.org/abs/2001.05084} {\bibinfo  {journal} {arXiv:2001.05084}\
  }\BibitemShut {NoStop}%
\bibitem [{\citenamefont {Su}\ and\ \citenamefont {Lin}()}]{su2020switching}%
  \BibitemOpen
\bibfield  {journal} {  }\bibfield  {author} {\bibinfo {author} {\bibfnamefont
  {Y.}~\bibnamefont {Su}}\ and\ \bibinfo {author} {\bibfnamefont {S.-Z.}\
  \bibnamefont {Lin}},\ }\href {https://arxiv.org/abs/2002.02611} {\bibinfo
  {journal} {arXiv:2002.02611}\ }\BibitemShut {NoStop}%
\bibitem [{\citenamefont {Kwan}\ \emph {et~al.}({\natexlab{a}})\citenamefont
  {Kwan}, \citenamefont {Hu}, \citenamefont {Simon},\ and\ \citenamefont
  {Parameswaran}}]{kwan2020exciton}%
  \BibitemOpen
\bibfield  {journal} {  }\bibfield  {author} {\bibinfo {author} {\bibfnamefont
  {Y.~H.}\ \bibnamefont {Kwan}}, \bibinfo {author} {\bibfnamefont
  {Y.}~\bibnamefont {Hu}}, \bibinfo {author} {\bibfnamefont {S.~H.}\
  \bibnamefont {Simon}}, \ and\ \bibinfo {author} {\bibfnamefont
  {S.}~\bibnamefont {Parameswaran}},\ }\href {https://arxiv.org/abs/2003.11560}
  {\bibfield  {journal} {\bibinfo  {journal} {arXiv:2003.11560}\ }
  ({\natexlab{a}})}\BibitemShut {NoStop}%
\bibitem [{\citenamefont {B{\"o}merich}\ \emph {et~al.}()\citenamefont
  {B{\"o}merich}, \citenamefont {Heinen},\ and\ \citenamefont
  {Rosch}}]{bomerich2020skyrmion}%
  \BibitemOpen
  \bibfield  {author} {\bibinfo {author} {\bibfnamefont {T.}~\bibnamefont
  {B{\"o}merich}}, \bibinfo {author} {\bibfnamefont {L.}~\bibnamefont
  {Heinen}}, \ and\ \bibinfo {author} {\bibfnamefont {A.}~\bibnamefont
  {Rosch}},\ }\href {https://arxiv.org/abs/2004.13684} {\bibinfo  {journal}
  {arXiv:2004.13684}\ }\BibitemShut {NoStop}%
\bibitem [{\citenamefont {Kwan}\ \emph {et~al.}({\natexlab{b}})\citenamefont
  {Kwan}, \citenamefont {Hu}, \citenamefont {Simon},\ and\ \citenamefont
  {Parameswaran}}]{kwan2020excitonic}%
  \BibitemOpen
\bibfield  {journal} {  }\bibfield  {author} {\bibinfo {author} {\bibfnamefont
  {Y.~H.}\ \bibnamefont {Kwan}}, \bibinfo {author} {\bibfnamefont
  {Y.}~\bibnamefont {Hu}}, \bibinfo {author} {\bibfnamefont {S.~H.}\
  \bibnamefont {Simon}}, \ and\ \bibinfo {author} {\bibfnamefont
  {S.}~\bibnamefont {Parameswaran}},\ }\href {https://arxiv.org/abs/2003.11559}
  {\bibfield  {journal} {\bibinfo  {journal} {arXiv:2003.11559}\ }
  ({\natexlab{b}})}\BibitemShut {NoStop}%
\bibitem [{\citenamefont {Zhang}\ and\ \citenamefont
  {Senthil}()}]{zhang2020quantum}%
  \BibitemOpen
  \bibfield  {author} {\bibinfo {author} {\bibfnamefont {Y.-H.}\ \bibnamefont
  {Zhang}}\ and\ \bibinfo {author} {\bibfnamefont {T.}~\bibnamefont
  {Senthil}},\ }\href {https://arxiv.org/abs/2003.13702} {\bibinfo  {journal}
  {arXiv:2003.13702}\ }\BibitemShut {NoStop}%
\bibitem [{\citenamefont {Stefanidis}\ and\ \citenamefont
  {Sodemann}()}]{stefanidis2020excitonic}%
  \BibitemOpen
\bibfield  {journal} {  }\bibfield  {author} {\bibinfo {author} {\bibfnamefont
  {N.}~\bibnamefont {Stefanidis}}\ and\ \bibinfo {author} {\bibfnamefont
  {I.}~\bibnamefont {Sodemann}},\ }\href {https://arxiv.org/abs/2004.03613}
  {\bibinfo  {journal} {arXiv:2004.03613}\ }\BibitemShut {NoStop}%
\bibitem [{\citenamefont {Nagaoka}(1966)}]{Nagaoka1966Ferro}%
  \BibitemOpen
\bibfield  {journal} {  }\bibfield  {author} {\bibinfo {author} {\bibfnamefont
  {Y.}~\bibnamefont {Nagaoka}},\ }\href {\doibase 10.1103/PhysRev.147.392}
  {\bibfield  {journal} {\bibinfo  {journal} {Phys. Rev.}\ }\textbf {\bibinfo
  {volume} {147}},\ \bibinfo {pages} {392} (\bibinfo {year}
  {1966})}\BibitemShut {NoStop}%
\bibitem [{\citenamefont {Lieb}(1989)}]{Lieb1989}%
  \BibitemOpen
  \bibfield  {author} {\bibinfo {author} {\bibfnamefont {E.~H.}\ \bibnamefont
  {Lieb}},\ }\href {\doibase 10.1103/PhysRevLett.62.1201} {\bibfield  {journal}
  {\bibinfo  {journal} {Phys. Rev. Lett.}\ }\textbf {\bibinfo {volume} {62}},\
  \bibinfo {pages} {1201} (\bibinfo {year} {1989})}\BibitemShut {NoStop}%
\bibitem [{\citenamefont {Mielke}(1992)}]{mielke1992exact}%
  \BibitemOpen
  \bibfield  {author} {\bibinfo {author} {\bibfnamefont {A.}~\bibnamefont
  {Mielke}},\ }\href
  {https://iopscience.iop.org/article/10.1088/0305-4470/25/16/011} {\bibfield
  {journal} {\bibinfo  {journal} {Journal of Physics A: Mathematical and
  General}\ }\textbf {\bibinfo {volume} {25}},\ \bibinfo {pages} {4335}
  (\bibinfo {year} {1992})}\BibitemShut {NoStop}%
\bibitem [{\citenamefont {Moon}\ \emph {et~al.}(1995)\citenamefont {Moon},
  \citenamefont {Mori}, \citenamefont {Yang}, \citenamefont {Girvin},
  \citenamefont {MacDonald}, \citenamefont {Zheng}, \citenamefont {Yoshioka},\
  and\ \citenamefont {Zhang}}]{Moon1995}%
  \BibitemOpen
  \bibfield  {author} {\bibinfo {author} {\bibfnamefont {K.}~\bibnamefont
  {Moon}}, \bibinfo {author} {\bibfnamefont {H.}~\bibnamefont {Mori}}, \bibinfo
  {author} {\bibfnamefont {K.}~\bibnamefont {Yang}}, \bibinfo {author}
  {\bibfnamefont {S.~M.}\ \bibnamefont {Girvin}}, \bibinfo {author}
  {\bibfnamefont {A.~H.}\ \bibnamefont {MacDonald}}, \bibinfo {author}
  {\bibfnamefont {L.}~\bibnamefont {Zheng}}, \bibinfo {author} {\bibfnamefont
  {D.}~\bibnamefont {Yoshioka}}, \ and\ \bibinfo {author} {\bibfnamefont
  {S.-C.}\ \bibnamefont {Zhang}},\ }\href {\doibase 10.1103/PhysRevB.51.5138}
  {\bibfield  {journal} {\bibinfo  {journal} {Phys. Rev. B}\ }\textbf {\bibinfo
  {volume} {51}},\ \bibinfo {pages} {5138} (\bibinfo {year}
  {1995})}\BibitemShut {NoStop}%
\bibitem [{\citenamefont {Roy}(2014)}]{Roy2014Geometry}%
  \BibitemOpen
  \bibfield  {author} {\bibinfo {author} {\bibfnamefont {R.}~\bibnamefont
  {Roy}},\ }\href {\doibase 10.1103/PhysRevB.90.165139} {\bibfield  {journal}
  {\bibinfo  {journal} {Phys. Rev. B}\ }\textbf {\bibinfo {volume} {90}},\
  \bibinfo {pages} {165139} (\bibinfo {year} {2014})}\BibitemShut {NoStop}%
\bibitem [{\citenamefont {Yang}\ \emph {et~al.}(2006)\citenamefont {Yang},
  \citenamefont {Das~Sarma},\ and\ \citenamefont {MacDonald}}]{Yang2006}%
  \BibitemOpen
  \bibfield  {author} {\bibinfo {author} {\bibfnamefont {K.}~\bibnamefont
  {Yang}}, \bibinfo {author} {\bibfnamefont {S.}~\bibnamefont {Das~Sarma}}, \
  and\ \bibinfo {author} {\bibfnamefont {A.~H.}\ \bibnamefont {MacDonald}},\
  }\href {\doibase 10.1103/PhysRevB.74.075423} {\bibfield  {journal} {\bibinfo
  {journal} {Phys. Rev. B}\ }\textbf {\bibinfo {volume} {74}},\ \bibinfo
  {pages} {075423} (\bibinfo {year} {2006})}\BibitemShut {NoStop}%
\bibitem [{\citenamefont {Chatterjee}\ \emph {et~al.}(2020)\citenamefont
  {Chatterjee}, \citenamefont {Bultinck},\ and\ \citenamefont
  {Zaletel}}]{Chatterjee2020Symm}%
  \BibitemOpen
  \bibfield  {author} {\bibinfo {author} {\bibfnamefont {S.}~\bibnamefont
  {Chatterjee}}, \bibinfo {author} {\bibfnamefont {N.}~\bibnamefont
  {Bultinck}}, \ and\ \bibinfo {author} {\bibfnamefont {M.~P.}\ \bibnamefont
  {Zaletel}},\ }\href {\doibase 10.1103/PhysRevB.101.165141} {\bibfield
  {journal} {\bibinfo  {journal} {Phys. Rev. B}\ }\textbf {\bibinfo {volume}
  {101}},\ \bibinfo {pages} {165141} (\bibinfo {year} {2020})}\BibitemShut
  {NoStop}%
\bibitem [{\citenamefont {Huang}\ \emph {et~al.}()\citenamefont {Huang},
  \citenamefont {Wang}, \citenamefont {Liu}, \citenamefont {Hu},\ and\
  \citenamefont {You}}]{huang2020quantum}%
  \BibitemOpen
  \bibfield  {author} {\bibinfo {author} {\bibfnamefont {X.-Y.}\ \bibnamefont
  {Huang}}, \bibinfo {author} {\bibfnamefont {T.}~\bibnamefont {Wang}},
  \bibinfo {author} {\bibfnamefont {S.}~\bibnamefont {Liu}}, \bibinfo {author}
  {\bibfnamefont {H.-Y.}\ \bibnamefont {Hu}}, \ and\ \bibinfo {author}
  {\bibfnamefont {Y.-Z.}\ \bibnamefont {You}},\ }\href
  {https://arxiv.org/abs/2005.01439} {\bibinfo  {journal} {arXiv:2005.01439}\
  }\BibitemShut {NoStop}%
\bibitem [{\citenamefont {Srivastava}\ and\ \citenamefont
  {Imamo\u{g}lu}(2015)}]{Srivastava2015}%
  \BibitemOpen
\bibfield  {journal} {  }\bibfield  {author} {\bibinfo {author} {\bibfnamefont
  {A.}~\bibnamefont {Srivastava}}\ and\ \bibinfo {author} {\bibfnamefont
  {A.}~\bibnamefont {Imamo\u{g}lu}},\ }\href {\doibase
  10.1103/PhysRevLett.115.166802} {\bibfield  {journal} {\bibinfo  {journal}
  {Phys. Rev. Lett.}\ }\textbf {\bibinfo {volume} {115}},\ \bibinfo {pages}
  {166802} (\bibinfo {year} {2015})}\BibitemShut {NoStop}%
\bibitem [{\citenamefont {Zhou}\ \emph {et~al.}(2015)\citenamefont {Zhou},
  \citenamefont {Shan}, \citenamefont {Yao},\ and\ \citenamefont
  {Xiao}}]{Zhou2015Berry}%
  \BibitemOpen
  \bibfield  {author} {\bibinfo {author} {\bibfnamefont {J.}~\bibnamefont
  {Zhou}}, \bibinfo {author} {\bibfnamefont {W.-Y.}\ \bibnamefont {Shan}},
  \bibinfo {author} {\bibfnamefont {W.}~\bibnamefont {Yao}}, \ and\ \bibinfo
  {author} {\bibfnamefont {D.}~\bibnamefont {Xiao}},\ }\href {\doibase
  10.1103/PhysRevLett.115.166803} {\bibfield  {journal} {\bibinfo  {journal}
  {Phys. Rev. Lett.}\ }\textbf {\bibinfo {volume} {115}},\ \bibinfo {pages}
  {166803} (\bibinfo {year} {2015})}\BibitemShut {NoStop}%
\bibitem [{\citenamefont {Peotta}\ and\ \citenamefont
  {T{\"o}rm{\"a}}(2015)}]{peotta2015superfluidity}%
  \BibitemOpen
  \bibfield  {author} {\bibinfo {author} {\bibfnamefont {S.}~\bibnamefont
  {Peotta}}\ and\ \bibinfo {author} {\bibfnamefont {P.}~\bibnamefont
  {T{\"o}rm{\"a}}},\ }\href {https://www.nature.com/articles/ncomms9944}
  {\bibfield  {journal} {\bibinfo  {journal} {Nat. Commun.}\ }\textbf {\bibinfo
  {volume} {6}},\ \bibinfo {pages} {8944} (\bibinfo {year} {2015})}\BibitemShut
  {NoStop}%
\bibitem [{\citenamefont {Hu}\ \emph {et~al.}(2019)\citenamefont {Hu},
  \citenamefont {Hyart}, \citenamefont {Pikulin},\ and\ \citenamefont
  {Rossi}}]{hu2019geometric}%
  \BibitemOpen
  \bibfield  {author} {\bibinfo {author} {\bibfnamefont {X.}~\bibnamefont
  {Hu}}, \bibinfo {author} {\bibfnamefont {T.}~\bibnamefont {Hyart}}, \bibinfo
  {author} {\bibfnamefont {D.~I.}\ \bibnamefont {Pikulin}}, \ and\ \bibinfo
  {author} {\bibfnamefont {E.}~\bibnamefont {Rossi}},\ }\href {\doibase
  10.1103/PhysRevLett.123.237002} {\bibfield  {journal} {\bibinfo  {journal}
  {Phys. Rev. Lett.}\ }\textbf {\bibinfo {volume} {123}},\ \bibinfo {pages}
  {237002} (\bibinfo {year} {2019})}\BibitemShut {NoStop}%
\bibitem [{\citenamefont {Julku}\ \emph {et~al.}(2020)\citenamefont {Julku},
  \citenamefont {Peltonen}, \citenamefont {Liang}, \citenamefont {Heikkil\"a},\
  and\ \citenamefont {T\"orm\"a}}]{julku2019superfluid}%
  \BibitemOpen
  \bibfield  {author} {\bibinfo {author} {\bibfnamefont {A.}~\bibnamefont
  {Julku}}, \bibinfo {author} {\bibfnamefont {T.~J.}\ \bibnamefont {Peltonen}},
  \bibinfo {author} {\bibfnamefont {L.}~\bibnamefont {Liang}}, \bibinfo
  {author} {\bibfnamefont {T.~T.}\ \bibnamefont {Heikkil\"a}}, \ and\ \bibinfo
  {author} {\bibfnamefont {P.}~\bibnamefont {T\"orm\"a}},\ }\href {\doibase
  10.1103/PhysRevB.101.060505} {\bibfield  {journal} {\bibinfo  {journal}
  {Phys. Rev. B}\ }\textbf {\bibinfo {volume} {101}},\ \bibinfo {pages}
  {060505(R)} (\bibinfo {year} {2020})}\BibitemShut {NoStop}%
\bibitem [{\citenamefont {Xie}\ \emph {et~al.}(2020)\citenamefont {Xie},
  \citenamefont {Song}, \citenamefont {Lian},\ and\ \citenamefont
  {Bernevig}}]{xie2019topology}%
  \BibitemOpen
  \bibfield  {author} {\bibinfo {author} {\bibfnamefont {F.}~\bibnamefont
  {Xie}}, \bibinfo {author} {\bibfnamefont {Z.}~\bibnamefont {Song}}, \bibinfo
  {author} {\bibfnamefont {B.}~\bibnamefont {Lian}}, \ and\ \bibinfo {author}
  {\bibfnamefont {B.~A.}\ \bibnamefont {Bernevig}},\ }\href {\doibase
  10.1103/PhysRevLett.124.167002} {\bibfield  {journal} {\bibinfo  {journal}
  {Phys. Rev. Lett.}\ }\textbf {\bibinfo {volume} {124}},\ \bibinfo {pages}
  {167002} (\bibinfo {year} {2020})}\BibitemShut {NoStop}%
\bibitem [{\citenamefont {Sondhi}\ \emph {et~al.}(1993)\citenamefont {Sondhi},
  \citenamefont {Karlhede}, \citenamefont {Kivelson},\ and\ \citenamefont
  {Rezayi}}]{SondhiSkyrmion}%
  \BibitemOpen
  \bibfield  {author} {\bibinfo {author} {\bibfnamefont {S.~L.}\ \bibnamefont
  {Sondhi}}, \bibinfo {author} {\bibfnamefont {A.}~\bibnamefont {Karlhede}},
  \bibinfo {author} {\bibfnamefont {S.~A.}\ \bibnamefont {Kivelson}}, \ and\
  \bibinfo {author} {\bibfnamefont {E.~H.}\ \bibnamefont {Rezayi}},\ }\href
  {\doibase 10.1103/PhysRevB.47.16419} {\bibfield  {journal} {\bibinfo
  {journal} {Phys. Rev. B}\ }\textbf {\bibinfo {volume} {47}},\ \bibinfo
  {pages} {16419} (\bibinfo {year} {1993})}\BibitemShut {NoStop}%
\bibitem [{\citenamefont {Wu}\ and\ \citenamefont
  {Das~Sarma}(2020{\natexlab{b}})}]{Wu2020Ferro}%
  \BibitemOpen
  \bibfield  {author} {\bibinfo {author} {\bibfnamefont {F.}~\bibnamefont
  {Wu}}\ and\ \bibinfo {author} {\bibfnamefont {S.}~\bibnamefont {Das~Sarma}},\
  }\href {\doibase 10.1103/PhysRevB.101.155149} {\bibfield  {journal} {\bibinfo
   {journal} {Phys. Rev. B}\ }\textbf {\bibinfo {volume} {101}},\ \bibinfo
  {pages} {155149} (\bibinfo {year} {2020}{\natexlab{b}})}\BibitemShut
  {NoStop}%
\end{thebibliography}%

\end{document}